\documentclass[conference,compsoc]{IEEEtran}

\usepackage{hyperref}
\IEEEoverridecommandlockouts

\usepackage{caption, subcaption, graphicx}
\usepackage{multirow, xspace, booktabs}
\usepackage[draft, inline, nomargin, index]{fixme}
\fxsetup{theme=colorsig,mode=multiuser,inlineface=\itshape,envface=\itshape}
\usepackage{svg}
\usepackage{amsmath}
\usepackage{float}
\usepackage{cleveref}
\usepackage{pifont}
\pdfoutput=1 

\crefformat{section}{\S#2#1#3}
\crefformat{subsection}{\S#2#1#3}
\crefformat{subsubsection}{\S#2#1#3}

\newcommand{\eg}{e.g.,\ }
\newcommand{\etal}{et al.\xspace}
\newcommand{\ie}{i.e.,\ }
\newcommand{\cf}{cf.\ }

\newcommand{\parait}[1]{\vspace{.05in}{\it \noindent #1 }}
\newcommand{\para}[1]{\vspace{.05in}{\bf \noindent #1 }}
\newcommand{\sellers}{\texttt{sellers.json}\xspace}
\newcommand{\ads}{\texttt{ads.txt}\xspace}

\newcommand{\misinfo}[1]{{\bf M}$_{\text{#1}}$\xspace}
\newcommand{\control}{{\bf C}$_{\text{ranked}}$\xspace}
\newcommand{\controlfull}{{\bf C}$_{\text{100K}}$\xspace}
\newcommand{\datastatic}[1]{{\bf D}$^{\text{static}}_{\text{#1}}$\xspace}
\newcommand{\datacrawls}{{\bf D}$^{\text{crawls}}$\xspace}
\newcommand{\databrands}{{\bf D}$^{\text{brands}}$\xspace}

\let\oldding\ding% Store old \ding in \oldding
\renewcommand{\ding}[2][1]{\scalebox{#1}{\oldding{#2}}}% Scale \oldding via optional argument

\newcommand{\bigvspace}{\vspace{-0.15in}}
\newcommand{\smallvspace}{\vspace{-0.1in}}

\usepackage{enumitem}

\usepackage{cite}
\usepackage{cleveref}
\usepackage{amsmath,amssymb,amsfonts}
\usepackage{algorithmic}
\usepackage{graphicx}
\usepackage{textcomp}
\usepackage{xcolor}
\usepackage{soul}
\usepackage{url}
\def\BibTeX{{\rm B\kern-.05em{\sc i\kern-.025em b}\kern-.08em
    T\kern-.1667em\lower.7ex\hbox{E}\kern-.125emX}}

\usepackage{breakurl}

\begin{document}

\thispagestyle{plain}
\pagestyle{plain} % Apply page number style to all pages

\title{The Inventory is Dark and Full of Misinformation: \\Understanding Ad Inventory Pooling in the Ad-Tech Supply Chain}

\author{
{\rm Yash Vekaria}\\
University of California, Davis
%yvekaria@ucdavis.edu
\and
{\rm Rishab Nithyanand}\\
University of Iowa
%rishab-nithyanand@uiowa.edu
\and
{\rm Zubair Shafiq}\\
University of California, Davis
%zubair@ucdavis.edu
} % end author

\maketitle

\begin{abstract}
Ad-tech enables publishers to programmatically sell their ad inventory to millions of demand partners through a complex supply chain. 
The complexity and opacity of the ad-tech supply chain can be exploited by low-quality publishers (e.g., misinformation websites) to deceptively monetize their ad inventory. 
To combat such deception, the ad-tech industry has developed transparency standards and brand safety products.
In this paper, we show that these developments still fall short of preventing deceptive monetization. 
Specifically, we focus on how publishers can exploit the ad-tech supply chain, subvert ad-tech transparency standards, and undermine brand safety protections by pooling their ad inventory with unrelated sites. 
This type of deception is referred to as ``dark pooling.''
Our study shows that dark pooling is commonly employed by misinformation publishers on various major ad exchanges, and allows misinformation publishers to deceptively sell their ad inventory to reputable brands.
Our work suggests the need for improved vetting of ad exchange supply partners, the adoption of new ad-tech transparency standards that enable end-to-end validation of the ad-tech supply chain, and the widespread deployment of independent audits like ours.
\end{abstract}
%\yash{Shouldn't we have email address in author names for people to contact us? And keywords for better indexing of the paper?}

%\begin{IEEEkeywords}
%Online advertising, Ad fraud, Dark pooling, Brand safety, Misinformation
%\end{IEEEkeywords}

\section{Introduction}

\para{The complexity of online advertising lends itself to fraud.}
A key to the success of online advertising is the ability of advertisers and publishers to programmatically buy and sell ad inventory across hundreds of millions of websites in real-time~\cite{yuan2013real}. 
Notably, Real-Time Bidding (RTB) allows publishers to list their ad inventory for auction at an ad exchange~\cite{openrtb-guidelines}. 
The ad exchange then asks its demand partners to bid on the ad inventory listed by its supply partners, based on the associated contextual and behavioral information.
The ad-tech supply chain is complex because it relies on hundreds of specialized entities to effectively buy and sell the ad inventory in real-time and at scale~\cite{lumascape}.
Adding to this complexity, each ad impression often gets sold and resold through multiple parallel or waterfall auctions~\cite{waterfall-bidding}. 
Such scale and complexity, combined with the opaque nature of the ad-tech supply chain, makes it a ripe target for fraud and abuse~\cite{10.1145/2420950.2420954,10.1145/2068816.2068843,10.1145/2594368.2594391,10.1145/2342356.2342394,10.1145/2663716.2663719,10.1145/2508859.2516682,7163024,10.1145/3317549.3323407,10.1145/3460120.3484547}. 
One of the most common types of ad fraud involves creating low-quality websites and monetizing their ad inventory.
Fraudsters attempt to drive large volumes of traffic to their website through various illicit means such as bots, underground marketplaces, traffic exchanges, or even driving legitimate traffic through click-bait and viral propaganda~\cite{10.1145/2815675.2815708,farooqi2017characterizing,10.1145/3447535.3462510}.
A notable example that motivated our work is that of the ``Macedonian fake news complex''~\cite{macedonian-fakenews,balkan-trump,fb-trump-election}. 
In this scheme, fraudsters created misinformation websites with misleading and clickbait headlines, aiming to go viral on social media, which led to tens of millions of monetized ad impressions.

\para{Advertisers are invested in preventing fraud.}
Ad-tech has safeguards to protect against this type of ad fraud by blocking the ad inventory of low-quality websites even when the ad impressions might be from legitimate users. 
Specifically, brand safety features supported by demand-side platforms aim to allow advertisers to block ad inventory of web pages that contain hardcore violence, hate speech, pornography, or other types of potentially objectionable content~\cite{iab-brand-safety}. 
All the effort of fraudsters would be wasted if they are unable to monetize their ad inventory through programmatic advertising due to these brand safety features. 
Fraudsters are known to exploit the opaque nature of the complex ad-tech supply chain to undermine brand safety protections by misrepresenting their ad inventory~\cite{bashir2019longitudinal}.
For example, in domain spoofing~\cite{domain-spoofing}, low-quality publishers mimic the URLs of reputable publishers in their ad inventory, thus deceiving reputable brands into purchasing their ad space even when their original domain is blocked due to brand safety concerns~\cite{domain-spoofing-threat,humansecurity,domain-spoofing-types}.
To combat ad fraud resulting from misrepresented ad inventory, the Interactive Advertising Bureau (IAB) introduced two transparency standards.
\ads~\cite{iab-adstxt} requires publishers to disclose all authorized sellers of their ad inventory.
\sellers~\cite{iab-sellersjson} requires ad exchanges to disclose all publishers and intermediate sellers involved in selling the ad inventory.
Together, when correctly implemented, these standards can reduce ad fraud by enabling buyers to verify the sources of the inventory they are purchasing.

\para{Transparency mechanisms to prevent fraud are falling short.}
There is increasing concern that the \ads and \sellers standards are either not widely adopted, implemented in ways that do not facilitate effective supply-chain validation, or intentionally subverted by malicious actors in a variety of ways.
In this paper, we empirically investigate these concerns. 
We find that the \ads and \sellers disclosures are plagued by a large number of compliance issues and misrepresentations. 
Most notably, we find extensive evidence of ``pooling'' of ad inventory from unrelated websites --- a practice known in the industry as ``dark pooling.'' 
This makes it impossible for a buyer to reliably identify the sources of the ad inventory (\ie where their ad will ultimately be placed). 
Dark pooling effectively enables low-quality publishers to ``launder'' their ad inventory, making it indistinguishable from that of well-reputed publishers.
To gain insight into how low-quality publishers might circumvent the transparency required by the \ads and \sellers standards, we selected a set of well-known misinformation websites as a case study. 
This choice is motivated by the known instances where ads from reputable brands have inadvertently ended up on such websites in the past~\cite{lee2021spillover, IASConsumerPerception, IASBrandFavorability, checkmyads, AdvertisersDemonetizedBreitbart, CVSBreitbart}.
Focusing on these misinformation websites, we confirm: (1) their widespread failure to comply with the \ads and \sellers standards; and (2) widespread prevalence of ad inventory pooling. 
We also find instances of reputable brands buying ad impressions on these misinformation websites, perhaps unintentionally. 
Taken together, we make three key contributions.

\parait{Measuring compliance with the transparency standards of \ads and \sellers.} 
We study a set of control and well-known misinformation websites to compare their compliance with \ads and \sellers. 
We find that although compliance issues are widespread even in the control set of websites, they are significantly more prevalent on misinformation websites.

\parait{Measuring the prevalence of (dark) pooling.} 
We measure the high prevalence of ad inventory pooling by our control and misinformation websites. 
By analyzing the \ads and \sellers files, we identified nearly 80 thousand instances of pooling.
We find that the misinformation pools are significantly more than twice as likely to pool ad inventory from unrelated websites than those that do not contain a misinformation website.
Upon further analysis of ad-related metadata in network traffic, we confirmed the use of 297 pools across 38 ad exchanges by misinformation websites.

\parait{Measuring the (in)effectiveness of brand safety tools.} 
We find ads from 55 reputable brands, including Forbes, GoDaddy, Harvard, Intel, Microsoft, Nike, Samsung, Tumblr, Yahoo!, Verizon, and Wayfair, on misinformation websites. 
We investigate the correlation between the prevalence of pooling and ads from reputable brands on misinformation websites. 
We find that misinformation websites that are part of at least one dark pool are nearly 20\% more likely to attract ads from reputable brands than those that are not part of a dark pool.
The responses to our disclosures indicate that reputable brands are generally unaware of their ads appearing on misinformation websites despite several using a brand safety service.

While there is some anecdotal evidence of a general lack of compliance with the ad-tech transparency standards and dark pooling \cite{DeepSeeDarkPoolSalesHouse, AdweekDarkPoolSalesHouse}, it does not systematically study these issues at scale. 
To the best of our knowledge, our work is the first to systematically study compliance with ad transparency standards and (dark) pooling at scale.
\section{Background} \label{sec:background}

In this section, we provide a high-level overview of the mechanisms behind the supply of programmatic ads (\Cref{sec:background:supply-chain}) and the 
vulnerabilities in the ad supply chain (\Cref{sec:background:vulnerabilities}).

\subsection{Programmatic advertising} 
\label{sec:background:supply-chain}
% What is the fundamental goal here: automated trading
%
Although there are a variety of mechanisms for programmatic advertising (\eg real-time bidding, header bidding, exchange bidding) and the participating organizations might differ, the types of entities involved in the supply
chain remain the same for each mechanism.

\begin{figure}[!t]
  \includegraphics[width=\linewidth]{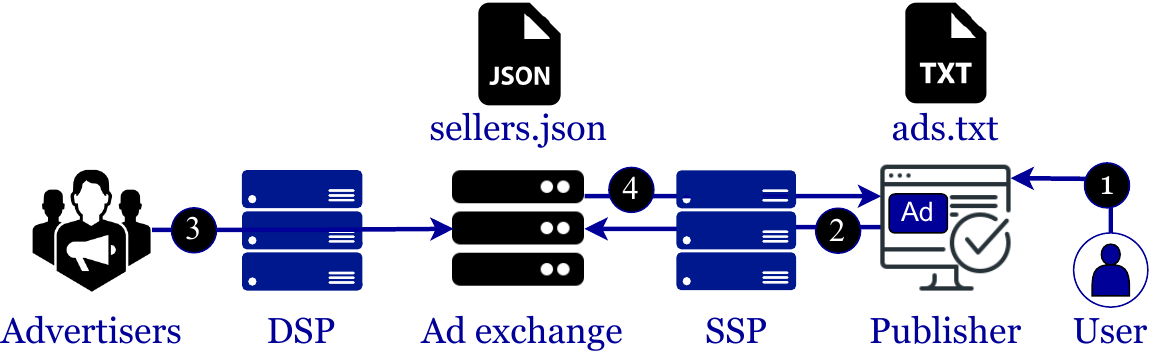}
  \caption{Programmatic advertising ecosystem: When a user visits a publisher website 
  (Step \ding[1.1]{182}), the publisher puts its ad-inventory for sale on ad exchanges via SSPs in 
  real-time (Step \ding[1.1]{183}). Advertisers bid for these slots via DSPs (Step \ding[1.1]{184}). Advertisement 
  of the winning bid is displayed to the user on the publisher website (Step \ding[1.1]{185}). To 
  mitigate fraud, advertisers use \texttt{sellers.json} of ad exchanges and 
  \texttt{ads.txt} of publishers to verify who is and who is not an authorized seller 
  of a given inventory.}
  \label{fig:adecosystem}
  \bigvspace
\end{figure}

\para{The programmatic advertising supply chain.}%
Programmatic advertising is made possible by the following entities illustrated in \Cref{fig:adecosystem}:
{\em supply-side platforms} (SSPs) for publishers to list their ad inventory in real-time, {\em ad exchanges} (AdX) which aggregate the inventory of multiple SSPs and facilitate bidding on individual ad slots, and {\em demand-side platforms} (DSPs) which allow advertisers and brands to identify targets for their ad creatives by suitably bidding on the inventory listed at ad exchanges. 
These entities work together to create a supply chain for ads as follows:
When a user visits a publisher, the ad inventory associated with that visit is put up for auction at an AdX by the SSP. 
DSPs, operating on behalf of advertisers and brands, then make bids on the ad inventory available at the AdX. 
These bids are informed by what is known (to the DSP) about the user and the publisher. 
The winner of the auction is then notified by the AdX and the associated ad creative is used to fill the ad slot on the publisher's website.

\para{Transparency in the supply chain.}
% Motivating the need for ads.txt and sellers.json
Crucial to the operation of the ad supply chain is that the participating 
organizations can trust that publishers and AdXs are not misrepresenting their inventories or their relationships with other entities. 
For example, DSPs need to confirm that the ad inventory that they are bidding on is actually associated with a particular publisher. 
Similarly, DSPs also need to confirm that the AdXs that they are purchasing ad inventory from are actually authorized to (re)sell that 
inventory. 
The absence of trust in this supply chain can lead to situations where DSPs
place premium bids for ad slots that are actually associated with non-premium publishers --- ultimately leading to a brand's ad creative appearing on websites that they may not want to be associated with.
% Introduce ads.txt and sellers.json
To foster trust and enable DSPs (Demand-Side Platforms) to perform basic verification of the ad inventory, the Interactive Advertising Bureau (IAB) introduced two standards: \ads and \sellers.

\para{The \ads standard.}
The \ads\footnote{``ads'' in \ads stands for Authorized Digital Sellers. Full specification of the \ads standard is available at:
\url{https://iabtechlab.com/wp-content/uploads/2021/03/ads.txt-1.0.3.pdf}} standard (introduced in 2017) aims to address ad inventory fraud by requiring each publisher domain to maintain an \ads file at the root level directory (\eg {\tt publisher.example/ads.txt}). 
The \ads file is supposed to contain entries for all AdXs that are authorized to sell or resell the ad inventory of the publisher. 
Each entry in the \ads file contains the following fields:
\begin{itemize}[leftmargin=.4cm]
    \item the authorized AdX, 
    \item the publisher ID assigned to the publisher domain within the AdX network, and 
    \item the authorized relationship between the publisher and authorized AdX --- \ie whether the AdX is authorized as a \texttt{DIRECT} seller or \texttt{RESELLER} of inventory for the domain.
\end{itemize}

%% How does it help
\noindent\textit{How \ads helps prevent fraud.}
When an ad request is sent by a publisher to an AdX (which issues bid requests to DSPs), the request contains the publisher ID and the domain associated with the inventory being listed.
Importantly, because publisher IDs are typically associated with an organization and not a domain, it is possible for multiple domains to share the same publisher ID. 
\ads enables verification that a website is not spoofing the domain in their ad requests. 
More specifically, \ads allows: 
\begin{itemize}[leftmargin=.4cm]
    \item AdXs to verify that the publisher ID in the ad request matches the publisher ID associated with the domain in the ad request and
    \item DSPs to verify that the AdX claiming to (re)sell the inventory of a domain is authorized by the domain to do so.
\end{itemize}
Before the \ads standard, there were no mechanisms to facilitate such checks and the sale of fraudulent inventory was widespread~\cite{bashir2019longitudinal}.

\para{The \sellers standard.}
Similar to the \ads standard, \sellers aims to mitigate ad inventory fraud and misrepresentation. The \sellers standard\footnote{Full specification of the \sellers standard is available at: \url{https://iabtechlab.com/wp-content/uploads/2019/07/Sellers.json_Final.pdf}} requires each AdX and SSP to maintain a \sellers file at the root level directory (\eg 
{\tt adx.example/sellers.json}).\footnote{We observed that several AdXs, including Google, use non-standard paths --- \eg Google's \sellers is located at \url{https://storage.googleapis.com/adx-rtb-dictionaries/sellers.json}} 
This \sellers file {\em must} contain an entry for each entity that may be paid for inventory purchased through the AdX --- \ie one entry for each partner that is an inventory source for the AdX. Each entry in the \sellers file contains the following fields:
\begin{itemize}[leftmargin=.4cm]
    \item the seller type which indicates whether the entry is associated with a \texttt{PUBLISHER}, an \texttt{INTERMEDIARY} (\ie inventory reseller AdX), or \texttt{BOTH} (\ie this entity has their own inventory and also resells other inventory);
    \item the seller ID associated with the inventory source (same as the publisher ID in \ads if this entry is associated with a publisher. From this point onwards we will refer to seller ID or publisher ID as seller ID); and
    \item the name and domain associated with the seller ID (these fields may be marked as ``confidential'' by AdXs to protect the privacy of publishers).
\end{itemize}

%% How does this help?
\noindent \textit{How \sellers helps prevent fraud.}
When a bid request is received by a DSP from an AdX that is compliant with the \sellers standard, it must contain information about the provenance of the inventory in a Supply Chain Object (SCO).\footnote{Supply Chain Object (SCO) contains an ordered list of all the entities involved in the ad transaction (\eg publisher $\rightarrow$ SSP $\rightarrow$ reseller $\rightarrow$ AdX).}
At a high level, the \sellers file provides a mechanism for DSPs to identify and verify all the entities listed in this SCO. This is done as follows:
\begin{itemize}[leftmargin=.4cm]
    \item When a bid request is received by the DSP, it should use the AdX's \sellers file to verify that the final AdX has an authorized relationship with the prior holder (an SSP or another AdX) of the inventory.
    \item The previous step is applied recursively (on all intermediate neighbors in the SCO) to verify the end-to-end authenticity of the inventory.
    \item The DSP then uses the \sellers files of all intermediaries and the \ads file of the publisher to verify that the publisher is legitimate and (re)sellers who handle the publisher's inventory are authorized to do so.
\end{itemize}
This capability for end-to-end validation of the SCO (Supply Chain Object) allows DSPs to identify instances where the ad inventory originates from low-quality publishers using fraudulent \ads files or is being sold by malicious intermediaries.

\subsection{Supply chain vulnerabilities} 
\label{sec:background:vulnerabilities}
Despite the introduction of the \ads and \sellers standards, there remain various vulnerabilities in the ad inventory supply chain.
Our investigation focuses on the vulnerabilities that enable low-quality publishers to monetize their ad inventory by misrepresenting or obscuring its source.
Some of these vulnerabilities arise from misrepresentations in the \ads and \sellers files, while others arise from pooling their low-quality inventory with the inventory of unrelated high-quality publishers. 
We refer to the former as {\em inventory misrepresentation} and the latter as {\em dark pooling}. 

\para{Inventory misrepresentation.}
Inventory misrepresentation arises from misrepresentations of ad inventory by publishers. 
It can be identified by discrepancies in the publisher's \ads file and is possible when DSPs and AdXs do not follow the \ads and \sellers standards.
Some examples of these misrepresentations include: 
\begin{itemize}[leftmargin=.4cm]
    \item a publisher's \ads file might incorrectly use seller IDs of other  publishers to suggest an authorized relationship with an AdX to boost the perception of its inventory. (Misrepresentations \#1 and \#2) 
    \item a publisher's \ads file might incorrectly indicate that a popular AdX is an authorized (re)seller of its inventory to boost its reputation with other AdXs. (Misrepresentation \#3) 
    \item a publisher's \ads file might have more than one entry of the same seller type for an AdX or \sellers files might associate a seller ID with multiple publishers or sellers making \ads and \sellers verification unreliable. (Misrepresentations \#4 and \#9)
    \item a publisher's \ads file might list authorized relationships with (re)sellers that do not have \sellers files, making end-to-end verification impossible. (Misrepresentation \#8) 
\end{itemize}

\para{Dark pooling.}
{\em Pooling} is a common strategy to share resources in online advertising. 
Consider, for example, the case where two or more publishers are owned by the same parent organization. 
In such scenarios, the ability to share advertising infrastructure and AdX accounts allows for more efficient operation and management. 
One way to identify the occurrence of pooling is by noting a single AdX-issued `seller ID' shared by multiple publisher websites. 
{\em Dark pools} are pools in which seller IDs are shared by organizationally-unrelated publishers (possibly of differing reputation). 
Note that ``dark pooling'' is a term of art that is commonly used in industry. 
While pooling is not itself a ``dark'' practice, pooling seller IDs of unrelated publishers is considered a ``dark'' practice because it deceives potential buyers about the actual source of the ad inventory~\cite{DeepSeeDarkPoolSalesHouse, AdweekDarkPoolSalesHouse}.

The seller ID defined in \ads and \sellers standards is also defined in the RTB protocol~\cite{GoogleOpenRTB, VungleOpenRTB}. 
Note that the payment after successful completion of an RTB auction is made to the publisher (i.e., the seller) associated with the seller ID~\cite{DoubleClickPublishers}.
Hence, it should be noted that simply using another domain's seller ID in ad requests from a website will result in any ad-related payments being made to the owner of the seller ID. 
Therefore, for revenue sharing, the creation of these pools needs to be facilitated either through intermediaries (e.g., SSPs) or by  collaboration between publishers.

\parait{End-to-end validation of pooled supply chains.}
Pooling leads to a break down of any brand or DSP's ability to perform end-to-end verification of the ad inventory supply chain. Specifically, the final step of verification highlighted in \Cref{sec:background:supply-chain} cannot be meaningfully completed unless {\em all} domains associated with a publisher's account are publicly known (and unfortunately, this is not the case). 
This is because the end-to-end verification of the ad inventory supply chain, as specified by the IAB, implicitly relies on trust that seller IDs are actually associated with specific organizations and that these associations are verified by AdXs. 
We illustrate this with an example.
\begin{itemize}[leftmargin=.4cm]
    \item Consider a publisher website {\tt sportsnews.example} which has a legitimate subsidiary: {\tt nbanews.example}. The publisher registers for an account with a popular AdX ({\tt adx}) and is issued the seller ID {\tt sellerid} after being vetted by {\tt adx}. It is expected that this website can now share this seller ID with its subsidiaries. Both websites will now list {\tt adx} as a {\tt DIRECT} seller through the {\tt sellerid} account in their \ads files. 
    \item The publisher now decides to share {\tt adx}-issued seller ID with {\tt fakesportsnews.example}, another sports news website but of low quality, for a cut of the revenue generated from ads shown on {\tt fakesportsnews.example}. In its \ads file, {\tt fakesportsnews.example} now adds {\tt adx} as a {\tt DIRECT} seller and also lists {\tt sellerid} as its seller ID. Note that {\tt fakesportsnews.example} would otherwise be unable to get directly listed on {\tt adx} and monetize its ad inventory due to its low quality. 
    \item When an ad request for some inventory is sent from {\tt fakesportsnews.example}, all basic supply chain validation checks are successful because the seller ID {\tt sellerid} is in fact registered by {\tt adx} in their \sellers file. Any bidding DSP will therefore operate under the assumption that the website receiving their ads has been vetted by {\tt adx} and is associated with {\tt sportsnews.example}.
    \item Complications only arise if the verifier notices that {\tt sellerid} was only registered to the owner of {\tt sportsnews.example} and the bid request actually originated at {\tt fakesportsnews.example}. However, invalidating the bid request simply because of this inconsistency will mean that even legitimate subsidiaries such as {\tt nbanews.example} cannot pool their inventory. Instead, additional checks are required to ascertain whether {\tt fakesportsnews.example} and {\tt sportsnews.example} are related or whether {\tt adx} vetted {\tt fakesportsnews.example} as well. This issue remains unaddressed by current validation mechanisms.
\end{itemize}

\begin{figure}[!t]
  \centering
  \includegraphics[width=\linewidth]{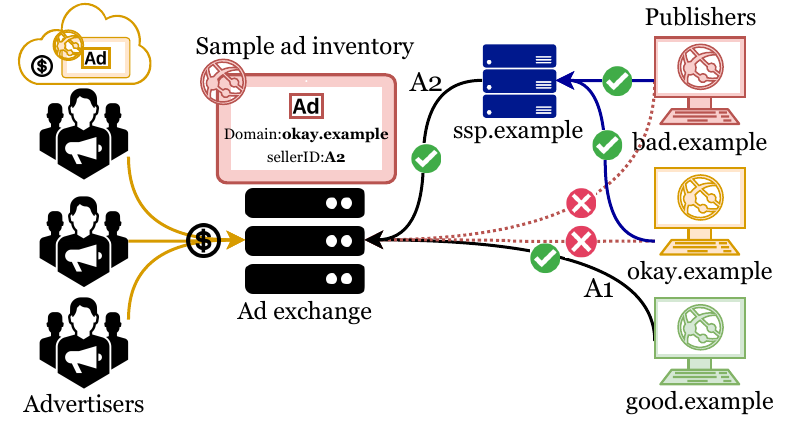}
  \caption{Illustration of pooling by an SSP --- A premium publisher (\texttt{good.example}) or an AdX-trusted intermediary SSP (\texttt{ssp.example}) can list on the AdX to obtain seller IDs {\tt A1} and {\tt A2} respectively. Whereas, a low-quality publisher (\texttt{bad.example}) or legitimate but unrecognized publisher (\texttt{okay.example}) are unable to directly list on the AdX. A legitimate publisher may not get listed on the AdX because, for instance, traffic requirements are not met. However, \texttt{bad.example} and \texttt{okay.example} are able to list on SSP, which essentially pools multiple publishers together. Bid request may misrepresent the inventory on \texttt{bad.example} as that of \texttt{okay.example} using the seller ID of the SSP (i.e., {\tt A2}). Reputable advertisers may bid on the inventory assuming that they are bidding on \texttt{okay.example}, when in fact their ad actually would end up on \texttt{bad.example}.} 
  \label{fig:pooling}
  \bigvspace
\end{figure}

\noindent\emph{Caveat.} The example described assumes collaboration between publishers --- {\tt sportsnews.example} and {\tt fakesportsnews.example}. This might be inadvertent in some cases --- \eg if {\tt sportsnews.example} and {\tt fakesportsnews.example} are both assigned the same seller ID through a common intermediary (an SSP, for example as shown in Figure~\ref{fig:pooling}). %\yash{Ref\#1}

In sum, by pooling various unrelated websites under a single seller ID, low-quality publishers can ``launder'' their ad inventory, rendering it indistinguishable from the inventory of high-quality publishers.
Moreover, this can occur when an AdX provides the seller ID to a trusted publisher (or an SSP), which then inadequately vets the low-quality publishers whose inventory it pools.
Figure~\ref{fig:pooling} illustrates this scenario of syndication-based pooling by some intermediary SSP. %\yash{Ref\#2}
As we show later, such pooling is common. 
In fact, we find some AdXs even providing services, via intermediaries, that facilitate pooling of unrelated entities. 

\section{Data} \label{sec:data}
In this section, we describe the selection of publishers that we study (\Cref{sec:data:websites})
and our methodology for collection of \ads, \sellers, and ad-related metadata associated with these websites
(\Cref{sec:data:methodology}).

\begin{figure*}[!t]
\bigvspace
\centering
  \includegraphics[width=2\columnwidth]{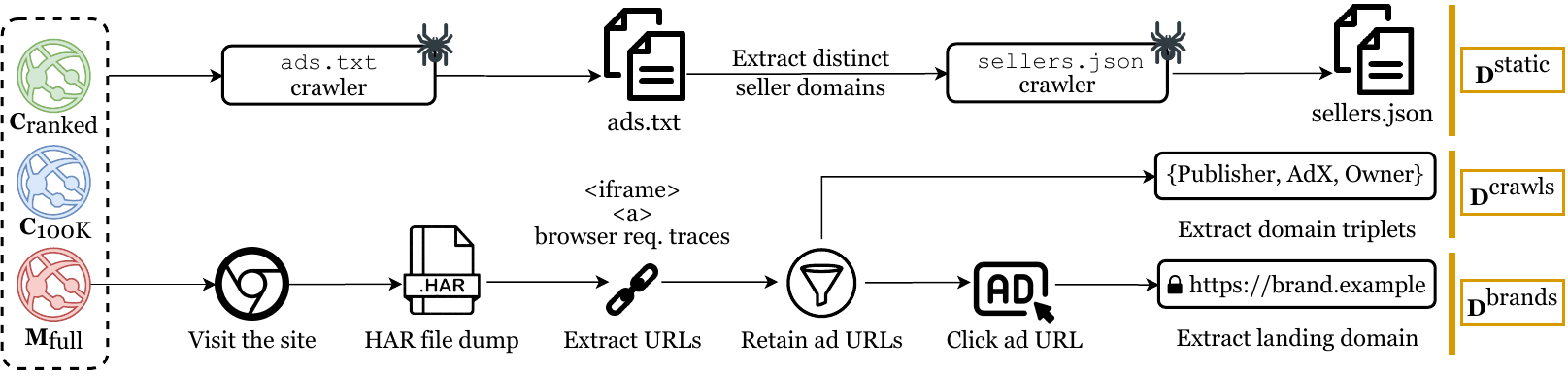}
  \caption{Overview of data collection methodology.}
  \label{fig:data:methodology}
  \bigvspace
\end{figure*}

\subsection{Publisher website selection} \label{sec:data:websites}
Our goal is to identify practices that hinder the end-to-end validation of the ad inventory supply chain, both among high-quality and low-quality websites. 
We use misinformation websites as a case study for low-quality websites and use comparably ranked websites from the Tranco list \cite{tranco} that have \ads as a stand-in for high-quality websites (referred to as a control).

\para{Selection of misinformation websites.}
% Misinformation generally refers to any misleading, deceptive, or false information --- including conspiracy theories, clickbait, hate content, pseudoscience, and satire. 
%
Since identifying misinformation websites is itself not the focus of our work, we leverage lists of misinformation websites curated in prior research by media scholars 
\cite{dataset1, dataset3} and computer scientists~\cite{hounsel2020identifying, zeng2020bad}.\footnote{For websites obtained from \cite{dataset3}, we discard those labeled as `state', `political', `credible', and `unknown'.} 
To construct our list of misinformation websites, we began by aggregating all websites from these lists and removing duplicates. 
This left us with 1276 websites. 
Next, we discarded 434 websites that were no longer functional. 
Finally, we additionally classified each misinformation website using multiple independent sources including Politifact, Snopes, MBFC, OpenSources, PropOrNot, and FakeNewsCodex to ensure that each remaining websites contained content that was undeniably misinformation. 
We excluded the websites that were now parked domains, seemed to have been repurposed, or had conflicting labels across different sources. 
This left us with a set of 669 {\em misinformation websites} (\misinfo{full}). 
Of these 669 websites, we created a subset of all the 251 websites that had an \ads file and were also present in the Tranco top-million list \cite{tranco} (\misinfo{ranked}). 
We use \misinfo{ranked} to compare the prevalence of \ads and \sellers discrepancies between misinformation and non-misinformation websites.

\para{Selection of benign (control) websites.}
To facilitate comparisons of the prevalence of compliance issues between misinformation and benign websites, we created a control set of non-misinformation websites (\control).
For each website in \misinfo{ranked}, we included the most similarly ranked non-misinformation website that also had an \ads file. 
We performed matching based on website domain ranks to avoid confounds related to website popularity. 
We also created a control set of the Tranco top-100K domains which contained an \ads file (\controlfull). 
This dataset was used to investigate the broad prevalence of pooling.
These four sets of websites (\misinfo{full}, \misinfo{ranked}, \control, and \controlfull) are the subject of our study.

\subsection{Data collection} \label{sec:data:methodology}
Our analysis relies on three sources of data: 
(1) \ads and \sellers files related to publishers, AdXs, and other intermediaries;
(2) bid/ad requests and responses during visits to a publisher domain; and 
(3) brands placing advertisements on a publisher domain.
An overview of our data collection is illustrated in \Cref{fig:data:methodology}. 
Table~\ref{tab:dataset_description} lists the notations for different datasets used throughout the paper.

\begin{table}[t]
    \centering
    \footnotesize
    \begin{tabular}{clc}
    \toprule
    \textbf{Notation}   &     \textbf{Description}      &       \textbf{Size}   \\
    \midrule
    \misinfo{full}  &   Complete set of misinformation domains studied  &   669 \\
    \misinfo{ranked}    &   Sites in \misinfo{full} with \ads \& part of Tranco-1M &   251 \\
    \control    &   Similar-ranked NM with \ads for each \misinfo{ranked}  &    251 \\
    \controlfull    &   Tranco Top-100K domains with \ads presence  &   20K \\
%    \datastatic{10/21}  &   \ads and \sellers crawled on 10/21  &   1.2K \\
    \datastatic{}  &   \ads and \sellers crawled on 02/22  &   1.4K \\
    \datacrawls &   (PD, AdX, OD) tuples from dynamic crawl of \misinfo{full}   &   2.8K    \\
    \databrands &   (PD, Brand) pairs from dynamic crawl of \misinfo{full}  &   4.2K    \\
    \bottomrule
    \end{tabular}
    \caption{{Description of dataset notations and sizes. NM represents non-misinformation websites. PD and OD represent publisher domain and owner domain respectively.}}
    \label{tab:dataset_description}
    \bigvspace
\end{table}

\para{\ads and \sellers files.} 
To build evidence for the occurrence of pooling and other misrepresentations, we need to analyze published \ads and \sellers files associated with publishers and ad-tech entities.

\parait{Processing \ads files.}
We searched for an \ads file at the root of each website in \misinfo{full}, \misinfo{ranked}, \control, 
and \controlfull. 
From these \ads files, we extracted the domains of all the entities that were listed as \texttt{DIRECT} sellers or \texttt{RESELLERS} of the publisher's inventory.

\parait{Processing \sellers files.}
For each seller identified in our \ads files, we crawled the \sellers file at the domain's root. 
When the \sellers file was unavailable at this path, a best-effort attempt was made to manually identify any non-standard location of this file. 
We manually searched for the \sellers for the top-1K ranked seller domains that were detected as {\tt INTERMEDIARY} or {\tt BOTH} and no \sellers was extracted by the crawler for that domain.
We performed a web search using ``$<$domain$>$ -- sellers.json'' query and looked for the JSON file on the official webpage of the seller. 
Two \sellers were detected in this manner -- google.com and pubmatic.com.
We then parsed each \sellers file to identify entities (and their domains) that were associated with \texttt{PUBLISHER}, \texttt{INTERMEDIARY}, or \texttt{BOTH} entries. 
Finally, until no new entities were discovered, we recursively fetched and parsed the \sellers file 
associated with the entities labeled as either {\tt INTERMEDIARY} or {\tt BOTH}. 
This recursive fetching ensures that we have complete coverage of all the supply chain entities that may sell the inventory of all publishers in our datasets.

\parait{The \datastatic{ } datasets.}
We crawled and processed \ads and \sellers files in February 2022. 
We refer to the dataset as \datastatic{}.
In total, \datastatic{} included over 98K relationships from \ads files and 2.4M relationships from \sellers files.

\parait{Limitations of this dataset.}
It should be noted that, by itself, this dataset {\em cannot present evidence that pooling is actually occurring.} 
This is because each publisher is responsible only for the content of their own \ads file, misrepresentation in other publishers' \ads files is not sufficient to imply pooling.

\para{Obtaining real-time bidding metadata.}
To identify concrete evidence of pooling, we constructed a dataset of real-time bidding metadata. 
These include bid requests, responses, redirects, and payloads associated with ad requests and responses. 
Seller ID is communicated in requests and responses to different entities in the advertising ecosystem. 
Therefore, during a crawl of a given publisher's website, observing an unrelated entity's seller ID in these metadata constitutes a more concrete evidence of pooling between them.

\parait{Crawling configuration.}
Following the best practices for crawling-based data collection~\cite{ahmad2020apophanies, jueckstock2021towards}, we collected this dataset using a web crawler driven by Selenium (v4.1.0) and the Chrome browser (v91.0) with bot mitigation strategies (multiple randomly timed full page scrolls and randomized mouse movements), Xvfb from a non-cloud vantage point, and a 30-second waiting time after the completion of each page load. 
Prior work has shown that the bidders and content of ad slots are impacted by previous browsing history \cite{cook2020inferring, musa2022atom}. 
Therefore, each page load was conducted with a new browser profile to avoid biases in our measurements of ad responses and content.
With these settings, we loaded each website twice in \misinfo{full} and saved the associated HTTP Archive (HAR) files and full-page screenshots.

\begin{comment}
\begin{figure}[h]
  % \centering
  \includegraphics[width=\columnwidth]{Plots/ad-request.pdf}
  \caption{Illustration of seller IDs matching in ad requests.
  %\yash{Any changes to this figure to improve it?}
  }
  \label{fig:ad-request}
\end{figure}
\end{comment}

\begin{figure}
     \centering
     \begin{subfigure}[b]{\columnwidth}
         \centering
         \includegraphics[width=\columnwidth]{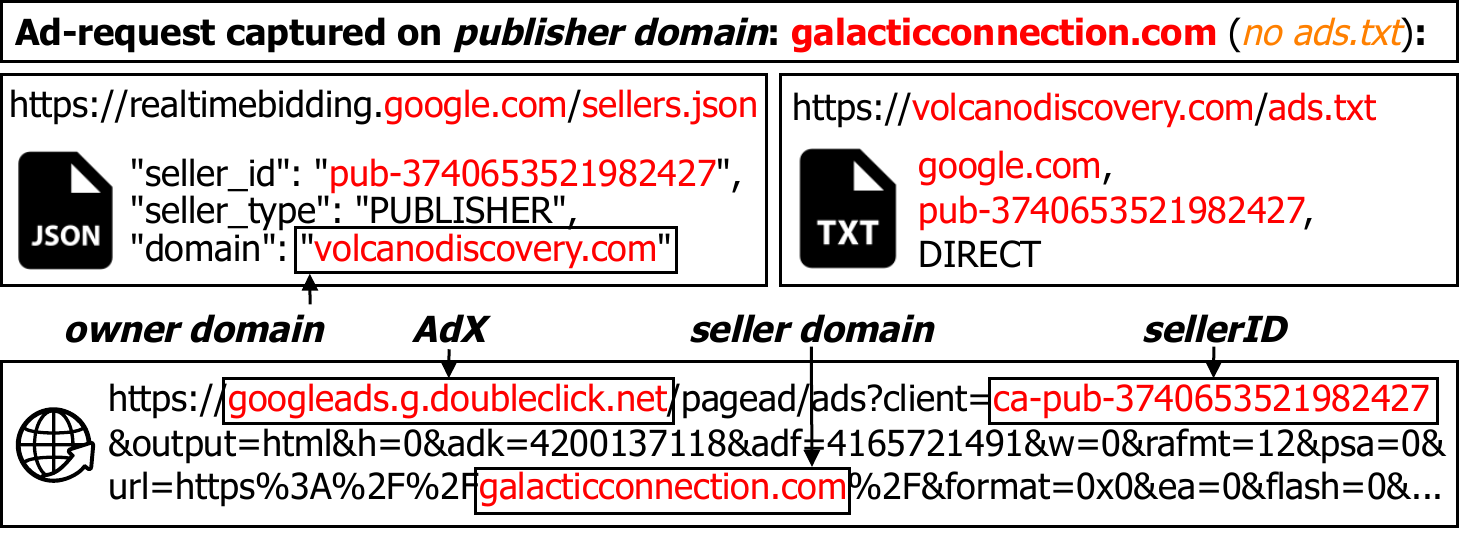}
         \caption{True positive case of ID matching an ad request}
         \vspace{0.175cm}
         \label{fig:ad-request-TP}
     \end{subfigure}
     \begin{subfigure}[b]{\columnwidth}
         \centering
         \includegraphics[width=\columnwidth]{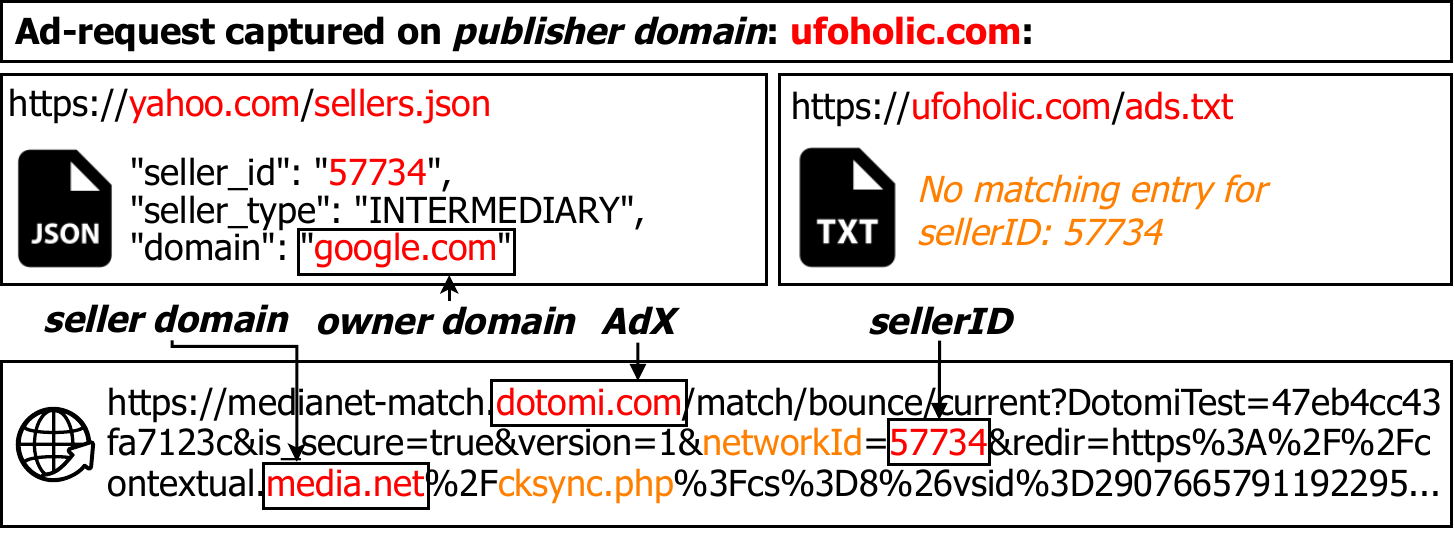}
         \caption{False positive case of ID matching in an ad request}
         \label{fig:ad-request-FP}
     \end{subfigure}
     \caption{Illustration of seller ID matching in ad requests for (a) true positive on the misinformation website: galacticconnection.com and (b) false positive on the misinformation website: ufoholic.com.}
     \bigvspace
\end{figure}

%\parait{Identifying ad-related requests and responses.} 
% Matching was performed using the {\it adblockparser} library.\footnote{\url{https://github.com/scrapinghub/adblockparser}} 

%% 
\parait{Extracting real-time bidding metadata from ad-related requests and responses.} 
From each HAR file, we first identified ad-related requests and responses 
by matching request URLs against well-known advertising filter lists used in prior research~\cite{iqbal2020adgraph}. 
We extracted the URLs, content, and HTTP {POST}-encoded data from each ad-related request and response. 
We then identified all (key, value) pairs using standard delimiters (e.g., \& in query parameters). 
% (including minor variations such as {\it key=``value''}). 
%
Finally, we matched the identified values with the seller IDs in from the \datastatic{} dataset. 
To mitigate false positives, we only matched ID strings with length greater than five characters.

Figure~\ref{fig:ad-request-TP} shows a sample ad request for doubleclick.net that is matched for the highlighted seller ID. 
Since volcanodiscovery.com listed in Google's \sellers and the misinformation website galacticconnection.com are unrelated, this represents a true positive instance of dark pooling.
%}
For each ad-related request and response, we identified the domain from which the request originated as the {\em publisher domain} (\ie observed inventory source) and the AdX domain owning the detected seller ID as the {\em AdX} (\ie inventory seller). 
We then used the \sellers of the {AdX/inventory seller} to identify the domain that owned the seller ID found in the ad request. 
This domain is labeled as the {\em owner domain} (\ie expected inventory source). 
The ({\em publisher domain}, {\em AdX}, {\em owner domain}) triples are used in later analysis.
Figure~\ref{fig:ad-request-FP} also shows a sample ad request on ufoholic.com, where one of the values matches with a Yahoo-issued seller ID that is owned by Google. 
However, Google-associated domains are absent from the ad request. 
The seller ID also does not exist in ufoholic.com's \ads. 
This match is deemed a false positive match and discarded from further analysis.

\parait{Methodology validation.}
We manually evaluated the accuracy of our method to extract metadata from ad-related requests. 
Specifically, we manually examined the requests and responses to verify that they did in fact include a $key$ that suggested that the
$value$ was associated with a seller ID. 
Our manual evaluation gave a false positive rate of 1.5\%.

\parait{The \datacrawls dataset.}
We label this dataset of ({\em publisher domain}, {\em AdX}, {\em owner domain}) triples
as \datacrawls. 
In total, the \datacrawls dataset consisted of 3.1K distinct triples observed across two crawls of 669 \misinfo{full} websites. 
In \Cref{sec:pooling}, we use these triples to determine (dark) pooling on misinformation websites.

\parait{Limitations of this dataset.} The programmatic advertising is auction-driven and participation from entities is non-deterministic. 
Therefore, any observations of entities and the IDs in requests and responses related to ads will vary from one crawl to the next, even when all other client-related factors are identical. 
Further, the browser provides a vantage point that typically only affords observations of the winners of real-time bidding auctions. 
Finally, it is possible that some communications regarding the involved seller ID are not visible to us due to hashing or other forms of obfuscation \cite{papadopoulos2017if}.
These limitations are unavoidable. 
It should be noted, however, that these limitations only impact the completeness of our findings and not the correctness.
In other words, the prevalence of pooling and other discrepancies, as measured by our crawls, are only a lower-bound for their actual prevalence.

\para{Identifying brands in advertisements.}
We also analyzed the brands whose ads appear on misinformation websites. 
To identify brands advertising on misinformation websites, we performed 10 separate crawls. 
This repetition was to account for the non-deterministic nature of programmatic advertising that results in a user receiving different ads on repeat visits to the same website. 
In each of the 10 crawls, after each page load was complete and the 30-second wait period ended, we clicked the DOM elements associated with each ad-related URL on the page. 
These clicks typically resulted in navigation to the brand's website.
We used this website's domain to label the brand associated with the ad.

\parait{Methodology validation.}
To test the effectiveness of this methodology, we conducted a pilot test on one crawl where we compared the brand
names identified through manual analysis and the automated approach. 
We found that in 30\% of the displayed ads, the automated approach failed to identify the brand associated with an ad. 
In these cases, failure was largely because some ad-related request URLs were associated with``unclickable'' elements of the ad. 
As a result, our automated approach could not trigger navigation to the brand's website.
To mitigate this issue, we supplemented our automated approach by manually annotating the ads on all crawls that could not be associated with a brand.
This process was relatively quick since most of the ads had been already automatically annotated with associated brands.

\parait{The \databrands dataset.}
We recorded all ({\em publisher, brand}) pairs identified with this methodology in \databrands dataset. 
In total, the \databrands dataset consisted of 4.2K distinct (publisher, brand) pairs and 2.1K unique brands.

\parait{Crawl success rate.}
Our crawling infrastructure for \ads and \sellers had a 100\% success rate (i.e., if a website had a file, we were able to crawl it without any failures). 
Dynamic web crawls did fail for a small percentage of websites ($<5\%$) due to timeouts. 
However, we were able to crawl all websites at least once since we performed multiple crawls for each website.

\parait{{Limitations.}}
We performed all crawls from one IP address, which could impact our analysis of brands. 
In other words, we might have observed more or less brands had we performed crawling from multiple IP addresses. 
%
%However, using a single IP-address could not have biased the measured advertisements in any other way given that we perform crawl randomization and we crawl each website independently by spawning up a fresh browser with cleared cookies/storage (aka ``stateless crawl'').
%}

\para{Ethical considerations.}
We discuss the ethics of our web crawling along three dimensions: infrastructure costs, privacy risks, and advertising costs caused by this study.
Overall, our study respects the principle of beneficence as outlined in the Menlo Report~\cite{MenloReport2012} and Belmont Report~\cite{BelmontReport1979} by maximizing the possible benefits and minimizing the harms.

\parait{Infrastructure costs.} Our crawls were used to measure the prevalence of compliance issues and misrepresentations. 
Our two dynamic crawls were not conducted concurrently to avoid stressing the web servers. 
Similarly, our additional static crawls for \ads and \sellers were performed six months apart. 
While our crawlers did not follow the {\tt robots.txt} directives (if present) on misinformation publishers, our crawling methodology is in line with ethical and legal considerations of such crawling-based auditing systems \cite{robotstxt,sandvig2014auditing,addicks2022van}. 
%and is not considered in violation of the CFAA per a supreme court ruling in {\em Van Buren vs. United States} \cite{addicks2022van}.
%\parait{Privacy risks.} By visiting only the front pages of each website using full-fledged web browsers, our dynamic crawls did not disobey the {\tt robots.txt} directives set by publishers. 
Also note that our study did not involve human subjects or gather any personal information.

\parait{Advertising costs.} To actually understand what brands are advertising on misinformation websites and what ad-exchange is responsible for showing that ad, we clicked on the ads shown during the page loads. The costs associated with our ad clicks are negligible (CPMs are in the order of fractions of cents and we clicked a total of 4247 ads). 
We believe these costs are justifiable given the benefit of understanding vulnerabilities in the ad-tech ecosystem.

%The three RQs
\section{Measuring Problematic Representations} \label{sec:pooling}
In this section, we answer the question: {\em what is the prevalence of pooling and other problematic representations on misinformation websites?} 
Specifically, we focus on measuring the prevalence of misrepresentations that hinder end-to-end supply chain validation. 
In \Cref{sec:pooling:misrepresentations}, we provide a broad overview of the types of misrepresentations commonly seen in \sellers and \ads files.
We compare the prevalence of these misrepresentations on control and misinformation websites.
In \Cref{sec:pooling:pooling}, we present evidence of ad inventory pooling and highlight cases of dark pooling by misinformation websites.

\subsection{Prevalence of misrepresentations} \label{sec:pooling:misrepresentations}
Certain types of misrepresentations in a publisher's \ads file or an AdX's \sellers file may prohibit automated end-to-end verification of the ad inventory supply chain.
We identify eight such problematic representations:
\begin{enumerate}[leftmargin=.4cm]
    
\item \emph{Misrepresented direct relationships:} The Publisher claims that an AdX is a {\tt DIRECT} seller of its inventory, but the AdX's \sellers lists it as an {\tt INTERMEDIARY} (reseller) relationship;

\item \emph{Misrepresented reseller relationships:} The Publisher claims that an AdX account is a {\tt RESELLER} of its inventory, but the AdX's \sellers associates the corresponding account as a {\tt PUBLISHER} (direct) entry;

\item \emph{Fabricated seller IDs:} A publisher's \ads claims that an AdX is authorized to sell its inventory via some seller ID, but the AdX's \sellers does not have {\em any} account associated with that specific ID;

\item \emph{Conflicting relationships:} A publisher claims the same type of relationship(s) with more than one seller ID on a given AdX in their \ads, but the AdX only lists one of these relationships in their \sellers;

\item \emph{Invalid seller type:} The \sellers does not use any of the three acceptable types ({\tt PUBLISHER}, {\tt INTERMEDIARY}, or {\tt BOTH}) to describe the source of the inventory associated with a specific seller ID;

\item \emph{Invalid domain names:} The \sellers does not present a valid domain name\footnote{While a buyer may still rely on the `name' field, it is not suitable for automated analysis because `name' is a free text field. Automated analysis is crucial as bid requests need to be programmatically validated in real-time and at scale.} in the `domain' field;

\item \emph{Confidential sellers:} The \sellers lists the domain associated with the seller ID as `confidential'. It should be noted that this is not a violation of the \sellers standard, but does prevent end-to-end supply chain verification because both the `domain' and `name' fields are redacted;

\item \emph{Intermediaries without \sellers:} An AdX's \sellers lists intermediaries that do not have a \sellers; and

\item \emph{Non-unique seller IDs:} The \sellers associates multiple publisher or seller domains with the same seller ID confounding the buyer's verification.

\end{enumerate}

\begin{table}[!t]
    \centering
    \footnotesize
    \begin{tabular}{clcc}
    \toprule
    \textbf{Index} & \textbf{Type}      &     \control      &       \misinfo{ranked}   \\
    \midrule
    1 & Misrepresented direct relationships         &    51\%   &   64\%    \\
    2 & Misrepresented reseller relationships       &    47\%   &   65\%    \\
    3 & Fabricated seller IDs             &    65\%   &   83\%    \\
    4 & Conflicting relationships                   &    33\%   &   49\%    \\
    \bottomrule
    \end{tabular}
    \caption{Prevalence of problematic representations in \ads from websites in \control and \misinfo{ranked}.}
    \label{tab:problematic:ads}
\end{table}

\begin{table}[!t]
    \centering
    \footnotesize
    \begin{tabular}{clcc}
    \toprule
       \textbf{Index} & \textbf{Type}      &     \textbf{No \misinfo{full}}      &       \textbf{$\geq$ 1 \misinfo{full}}   \\
    \midrule
        5 & Invalid seller type   &   0.7\%         &     0\%     \\
        6 & Invalid domain names   &   0.8\%         &     54.8\%  \\
        7 & Confidential sellers  &   0.1\%         &     46.1\%  \\    
        8 & Intermediaries w/o \sellers    &   13.3\%        &     49.8\%  \\
        9 & Non-unique seller IDs      &   62.6\%        &     95.3\%  \\
    \bottomrule
    \end{tabular}
    \caption{Fraction of \sellers entries that contain different problematic representations from AdXs serving no \misinfo{full} websites and at least one \misinfo{full} website.}
    \label{tab:problematic:sellers}
    \bigvspace
\end{table}

\Cref{tab:problematic:ads} compares the prevalence of misrepresentations in \ads files of \control and \misinfo{ranked} websites. 
We find a statistically significant difference in the number of errors present in \ads files from \control and \misinfo{ranked} websites ($\chi^2$-test; $p < .05$). 
We find that misinformation websites are more likely to contain higher rates of \ads misrepresentations that result in failed supply chain validation, even when controlling for website rank.
\Cref{tab:problematic:sellers} compares the prevalence of misrepresentations in \sellers of AdXs that serve \misinfo{full} 
(344 AdXs) websites with the \sellers from AdXs that do not serve any of our \misinfo{full} websites (483 AdXs). 
Again, we see that the AdXs that engage with misinformation websites are significantly more likely to have misrepresentations in their \sellers that result in the inability to perform supply chain validation. 
Taken together, our results highlight the lack of compliance with \ads and \sellers standards and their current inability to allow end-to-end supply chain validation. 
This problem is especially pronounced for the ad inventory of misinformation publishers.

\subsection{Prevalence of pooling} \label{sec:pooling:pooling}
As described in \Cref{sec:background:vulnerabilities}, pooling is the practice of using a single AdX account to manage the inventory of multiple websites. %\yash{In section 2.2, we describe pub-pub pooling and ssp-pooling, but here we just refer pooling to be ssp-based, is that fine or confusing?}
This results in a single AdX-issued seller ID being associated with multiple websites. 
Although this practice enables more efficient management of advertising resources for publishers, it comes at the cost of increased opacity in the advertising ecosystem and reduces the effectiveness of the end-to-end supply chain validation mechanisms.

\para{Gathering evidence of pooling with the \datastatic{} dataset.}
We begin by identifying evidence of pooling in the \controlfull and \misinfo{full} websites from our \datastatic{} dataset. 
We use this dataset of \ads files associated with the Tranco top-100K domains to identify all cases where multiple domains listed the same seller ID and AdX as a seller of their inventory. 
In total, we observed 79K unique pools --- \ie 79K unique (seller ID, AdX) pairs were observed to have been shared by multiple publisher domains. 
Of these 79K pools, 11\% (8.7K) also included at least one of the misinformation websites in \misinfo{full}. 
We refer to these 79K pools identified through the \datastatic{} dataset as {\em static pools}.
The size of these pools ranged from 2 to nearly 9K domains, with an average of 70 domains per pool.

\para{Characteristics of pools identified in the \datastatic{} dataset.}
These above-reported pool sizes were certainly larger than what we anticipated and necessitated additional inspection for a better understanding of our findings. 
Specifically, we paid attention to the organizational relationships between pooled entities and whether pooling was occurring due to some ad-tech related mechanism.

\parait{Organizational homogeneity of pools.} 
From a cursory manual inspection of our pools, we observed (rather unsurprisingly) that larger pools appeared to contain many organizationally unrelated domains --- \ie they were {\em heterogeneous}. 
To measure the prevalence of such types of pools at scale, we mapped each domain in a pool to their parent organization using the DuckDuckGo entity list \cite{DDG-trackerradar} and labeled each pool as follows:
\begin{enumerate}
 \item{\em Homogeneous:} Pools whose member domains could all be mapped to a single parent organization;
 \item{\em Potentially homogeneous:} Pools for which the parent organizations of all domains could not be identified. However, all domains that could be mapped were found to have the same parent organization;
 \item{\em Heterogeneous:} Pools whose member domains were owned by more than one parent organization; and 
 \item{\em Unknown:} Pools for which no domain could be mapped to a single parent organization.
\end{enumerate}

\begin{figure}[!t]
  \centering
  \includegraphics[width=\linewidth]{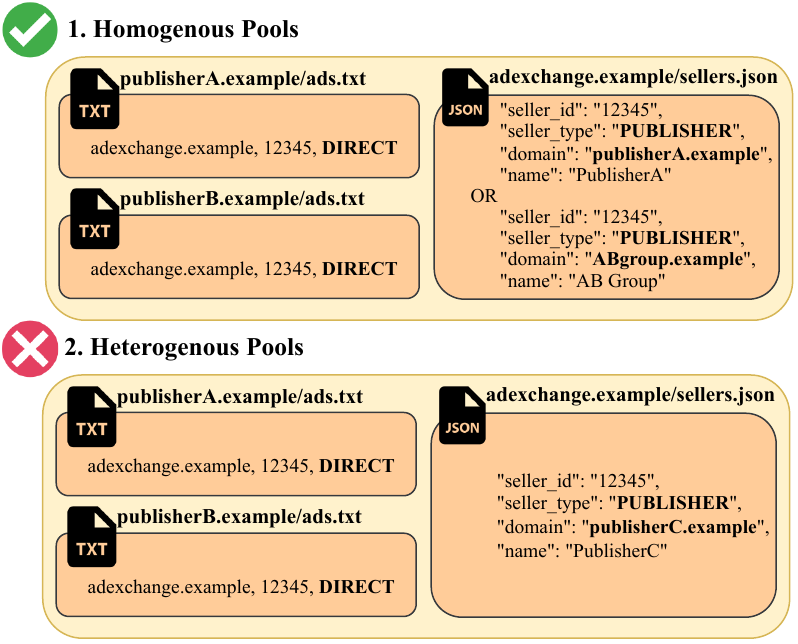}
  \caption{Categorization of pools based on the relation of the publishers with the domain owner organization. In the homogeneous pool, \textit{PublisherA} and \textit{PublisherB} authorize seller account 12345 on \textit{adexchange.example} as their direct seller. The \sellers of \textit{adexchange.example} recognizing 12345 as an account owned by either \textit{PublisherA}, \textit{PublisherB}, or \textit{AB Group} represent all valid cases of pooling assuming \textit{PublisherA} and \textit{PublisherB} are related (in this case owned or operated by \textit{AB Group}). 
  If the \sellers of \textit{adexchange.example} shows that the seller account 12345 is owned by \textit{PublisherC} and \textit{PublisherA} or \textit{PublisherB} are unrelated to \textit{PublisherC}, then this represents a case of heterogeneous pool, which we consider a dark pool.}
  \label{fig:dark_pooling_classification}
  \bigvspace
\end{figure}

\Cref{fig:dark_pooling_classification} illustrates how pools are categorized into homogeneous and heterogeneous.
\Cref{tab:pooling:homogeneity} provides a breakdown of the prevalence of different types of pools. 
We make three key observations. 
First, we notice that {\em heterogeneous pools comprise a large fraction of all pools} --- a deviation from the expectation that pools are allowed in order to facilitate resource sharing between sibling domains.
The high incidence rates of heterogeneous pools in non-misinformation websites also \textit{suggests} that there may be legitimate (\ie not ill-intentioned) mechanisms that facilitate seller ID sharing between organizations.
Second, {\em pools containing misinformation websites are statistically significantly more likely to be heterogeneous} (85\%) than pools without misinformation websites (41\%) [$\chi^2$-test; $p < .05$]. 
Finally, we see that {\em pools containing misinformation websites are statistically significantly larger} (412.1 websites/pool) than pools without misinformation websites (20.3 websites/pool) [2-sample $t$-test: $p < .05$; {$u$-test: $p < .05$}]. 
Taken together, the latter two findings lend credence to the thesis that misinformation websites are effectively ``laundering'' their ad inventory by participating in mechanisms that facilitate large heterogeneous pools.

\begin{table}[th]
\footnotesize
\begin{tabular}{p{.85in}lp{.2in}lp{.2in}}
\toprule
{}  & \multicolumn{2}{c}{{\bf Pools w/ \misinfo{full}}}  & \multicolumn{2}{c}{{\bf Pools w/o \misinfo{full}}}\\\cline{2-5}
{\bf Pool Type} \\ [-1em] & {\# Pools} & {$\mu_{size}$}    & {\# Pools}  & {$\mu_{size}$} \\
\midrule
{Homogeneous}                & 40 (0.4\%)    & 2.6          & 6.7K (9.6\%)   & 2.6          \\ 
{Po. Homogeneous}           & 913 (9.1\%)   & 18.8         & 18.4K (26.6\%) & 7.0          \\ 
{Heterogeneous}              & 8.6K (85.0\%) & 482.5        & 28.4K (41.0\%) & 42.2         \\ 
{Unknown}                    & 563 (5.6\%)   & 4.3          & 15.7K (22.7\%) & 3.9          \\ 
\midrule
All pools                    & 8.7K          & 412.1        & 70.5K          & 20.3         \\
\bottomrule
\end{tabular}
\caption{\textbf{Prevalence of pools from \datastatic{} in \controlfull.} Pools are broken down by organization homogeneity and whether they contained
a misinformation website from the \misinfo{full} dataset. $\mu_{size}$ denotes the average (mean) number of websites in a 
pool.}
\label{tab:pooling:homogeneity}
\bigvspace
\end{table}

\parait{Pools facilitated by authorized ad-tech mechanisms.} 
Our findings about the high rate of heterogeneous pools of large sizes, even among non-misinformation websites, suggest that there are ad-tech mechanisms that organically facilitate pooling.
After further investigation we found that many of the heavily pooled (seller ID, AdX) pairs appeared to be issued by a small number of AdXs whose \sellers file indicated that the issued seller IDs were not associated with specific publishers but instead other ad platforms (AdXs or SSPs). 
In other words, the seller ID issuing AdX's \sellers file indicated that the `owner domain' of the pooled seller ID was another AdX/SSP --- suggesting that these pooling mechanisms might be authorized by the AdX platforms themselves for aggregating and reselling ad inventory of different publishers. 
\Cref{tab:pooled_domains} shows that three of the most commonly pooled owner domains belong to large AdXs ({\tt google.com}, {\tt justpremium.com} owned by GumGum, and {\tt townnews.com}). 
Most notably, nearly 25\% and 12\% of the pools that used GumGum- and Google-owned seller IDs also contained known misinformation websites. 
For example, {\tt 100percentfedup.com}, a website that promoted anti-vax and stolen-election theories, received ads through pools using 
Google-owned seller IDs issued by the AdX `Index Exchange'. 
In contrast, TownNews, an advertising firm focused on serving local media organizations did not have a single pool containing known 
misinformation websites.

To investigate the prevalence of pooling, we looked for AdX-sanctioned programs that might require pooling --- \ie is there public documentation of {\em authorized} programs to allow unrelated publishers to pool their inventory through intermediaries. 
Notably, we found public documentation of Google's Multiple Customer Management (MCM) program that allows `Google MCM-partner' organizations to manage the inventory of multiple client publishers through a single account \cite{GoogleMCM}. 
As a result, all the publishers that are managed by an MCM partner are served ads via the same seller ID of the intermediary MCM organization. 
Our results show that misinformation websites are able to monetize their ad inventory by being part of these MCM networks.
{\em Our results highlight a violation of Google's own policies regarding advertising on websites `making unreliable claims' or `distributing manipulated media'} \cite{google-ad-guidelines}. 
However, public documentation does not clearly state whether Google delegates all website and content verification responsibilities to their MCM partners and therefore it remains unclear if the violation is a failure of Google's own verification practices or those of their MCM partners. 
Similarly, the pooled misinformation websites using GumGum-owned seller IDs were also in violation of GumGum's content policy \cite{gumgum-policy}.

\begin{table}[th]
    \centering
    \footnotesize
    \begin{tabular}{p{0.95in} l c c}
        \toprule
         \textbf{Type} & {\bf Domain} & {\bf Pools} & {\bf Pools w/ \misinfo{full}}  \\\midrule
         \multirow{5}{*}{Owner of sellerID} & {google.com}          & 5.1K       & 598 \\
         &{gannett.com}          & 370       & 5 \\
         &{justpremium.com}          & 337       & 84 \\ 
         &{townnews.com}          & 313       & 0 \\
         &{hearst.com}          & 219       & 1 \\
         \midrule
         \multirow{5}{*}{AdX issuer of sellerID} &          {google.com}          & 10.3K       & 461 \\
         & {taboola.com}          & 6.6K       & 132 \\
         & {freewheel.com}          & 3.9K        & 625 \\
         & {pubmine.com}          & 3.6K        & 2 \\
         & {openx.com}          & 2.4K        & 524 \\
         \bottomrule
    \end{tabular}
    \caption{\textbf{Most pooled domains and AdXs from \datastatic{}.} 
    The top five rows represent the most frequently observed domains whose seller IDs were used in pools. 
    The bottom five rows represent the most frequently observed AdXs who issued the seller IDs that were used for pooling.}%\yash{We don't have space to add similar table for dynamic crawls}}
    \label{tab:pooled_domains}
    \smallvspace
\end{table}

\parait{Pools using seller IDs with hidden or unknown owner domains.}
During our investigation, we also discovered that many AdX's \sellers files did not allow identification of the owner domain of the seller ID that was used. 
This comprised nearly half of all identified pools. 
The breakdown of reasons for this is provided in \Cref{tab:static:reasons_failed_owner_id}.
Here, we see that the most common reasons for failed identification of the owners of seller IDs being used in pooling are: (1) the seller ID is
itself unlisted in the issuing AdX's \sellers file and (2) the unavailability of a public \sellers from the owner domain that owned the AdX-issued seller ID (when owner domain is not a PUBLISHER type entry).
It is important to note that any of the reasons shown in \Cref{tab:static:reasons_failed_owner_id} would result in the impossibility of any end-to-end supply chain verification.
Interestingly, we find no statistical differences ($\chi^2$-test; $p < .05$) between the reasons for failed identification of owners of non-misinformation and misinformation pools.
This suggests that the issues of poor compliance with end-to-end supply chain verification procedures are industry-wide and no specific cause for these failures is exploited by misinformation websites.

\begin{table}[th]
\centering
\footnotesize
\begin{tabular}{p{1.5in}cc}
\toprule
{\bf Reason}    &   \textbf{All pools}   &  \textbf{Pools w/ \misinfo{full}} \\
\midrule
Total pools                 &   79K         &   8.7K    \\
\midrule
seller ID unlisted              &   20.9K       &   2.5K    \\
\sellers not public         &   16.5K       &   2.0K    \\
Owner not listed            &  2.6K         &   135     \\
Owner is \textit{confidential}     &   3.4K        &   86  \\
\bottomrule
\end{tabular}
\caption{{\bf Pools from \datastatic{} using IDs of unknown owners.} Reasons for failed identification of the owners of seller IDs used in pools.}
\label{tab:static:reasons_failed_owner_id}
\smallvspace
\end{table}

\begin{figure}[!t]
\bigvspace
 \centering
  \includegraphics[width=.97\columnwidth]{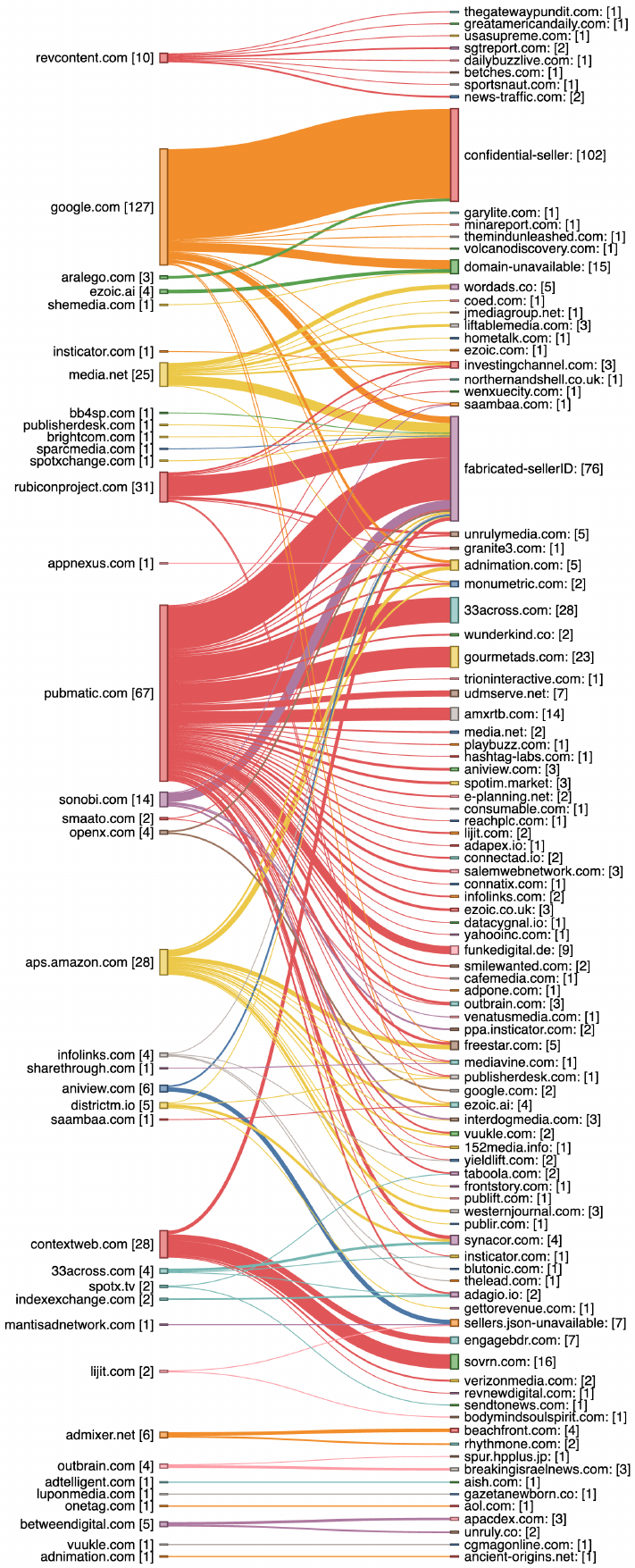}
  \vspace{.09in}
  \caption{Dark pooling relationships between AdXs (left) and owner domains of AdX-issued pooled IDs (right) for 297 unique pools observed during the crawls of \misinfo{full} websites. The counts represent the number of distinct misinformation websites pooled.}
  \label{fig:dynamicDarkSankey}
  \bigvspace
\end{figure}

\para{Finding occurrences of pooling with the \datacrawls dataset.}
Because of the high rates of misrepresentations, unreliability of publisher-sourced \ads files, and the incompleteness of AdX-sourced \sellers files, it is important to note that our analysis of the \datastatic{} can only be used as evidence that suggests the widespread practice of \textit{potential} dark pooling. 
In order to confirm a dark pool's existence with certainty we need to observe it in a live page load.
To this end, we leverage the set of all (publisher domain, AdX, owner domain) triples recorded in our \datacrawls dataset (\cf \Cref{sec:data:methodology}). 
Since these were obtained from actual ad-related metadata from crawls of the \misinfo{full} dataset, they provide concrete evidence of pooling actually being leveraged by known misinformation websites (\ie dark pooling).
In total, we gathered 2.8K (publisher domain, AdX, owner domain) triplets through two crawls of \misinfo{full} websites from which we identified 297 pools across 38 ad exchanges.
These 297 pools are depicted in Figure~\ref{fig:dynamicDarkSankey}
Of these, 218 pools (73.4\%) overlapped with those identified in our analysis of the \datastatic{} dataset and 79 were new. 
The non-existence of 79 pools in the \datastatic{} dataset prevented us from classifying them and this once again highlights the ad industry's poor compliance with \ads and \sellers standards.

Google and PubMatic were found to be the issuers of the seller IDs associated with 120 and 48 pools, respectively. 
These pools enabled advertising supply chains for 127 (Google) and 67 (PubMatic) misinformation websites.
33Across and Gourmet Ads were found to be the owners of seller IDs that were shared by the most number of misinformation websites (28 and 23 websites, respectively). 
Both seller IDs were issued by PubMatic.
Other notable AdXs (and count of the number of seller IDs issued by them which were pooled by misinformation websites) include Rubicon Project (now Magnite) (34), ContextWeb (now PulsePoint) (30), Amazon (28), and media.net (25).

\parait{Homogeneity of \datacrawls pools.}
From 297 {distinct} pools, we were able to identify the presence of {15} homogeneous and {203} heterogeneous pools. The homogeneity of the remaining pools could not be determined.
The largest homogeneous pool shared a seller ID issued to {\tt funkedigital.de} by PubMatic. This pool included nine websites such as {\tt principia-scientific.org}, 
{\tt allnewspipeline.com}, {\tt russia-insider.com} --- Media Bias/Fact Check identified all the nine websites as `Conspiracy Theory' 
or `Propaganda' related with `Low' factual reporting and having `Right' to `Extreme-Right' bias. We identified stories related to climate change denial, vaccination misinformation, 
and pro-insurrection views --- all in violation of PubMatic's own content guidelines for publishers \cite{pubmatic-policy}. 
Incidentally, a seller ID on Pubmatic was also associated with the largest heterogeneous pool with 47 unique misinformation websites, including {\tt drudgereport.com} and {\tt worldtruth.tv}. 
Unfortunately, PubMatic's \sellers file did not list the seller ID associated with this heterogeneous pool, suggesting that it was employing fabricated or unlisted ID for pooling.

\parait{\datacrawls pools and the Google MCM program.}
In order to identify occurrences of pooling in Google's MCM program, we identified pools associated with the seller IDs issued by Google to MCM partners.
Of the {203} {unique} heterogeneous pools identified, a vast majority were labeled as confidential in Google's \sellers \cite{PubIDBlocking} but we were able to link {15} to Google's MCM program based on public documentation. 
In total, Google's MCM partners were associated with 27 misinformation websites. %, 6 of which via Google and remaining via other AdXes. 
Some of these MCM partners whose Google-issued seller IDs were pooled by misinformation websites include Adnimation, Ezoic, etc.
Misinformation websites supported by Google's MCM program included {\tt 369news.net} (pseudoscience or anti-vaxx theories) and {\tt truthandaction.org} (extreme-right propaganda and/or misinformation), amongst other similar websites. 
The MCM partners most frequently found to be using their Google-issued seller ID for pools containing misinformation websites were Monumetric (5 pools) and Freestar (4 pools).

\begin{comment}
    Tranco Rank	& Avg.DynamicPoolCounts
    0-1K	2.5
    1K-5K	4.7
    5K-10K	6.67
    10K-50K	6.4
    50K-100K	5.75
    100K-200K	4
    200K-300K	3.4
    300K-400K	2.5
    400K-500K	1.67
    500K-600K	2.8
    600K-700K	8.33
    700K-800K	2
    800K-900K	2
    900K-1M	4.85
    1M+ Known	3.2
    Unranked	2.03
\end{comment}

\para{Takeaways.}
Our analysis shows a widespread failure to adhere to the \ads and \sellers standards and the compliance is even more weaker 
amongst misinformation websites (\Cref{sec:pooling:misrepresentations}). 
This poor adherence has one major consequence: end-to-end validation of the ad-inventory supply chain is not always
possible, particularly in the case of misinformation websites. 
Further compounding supply chain validation challenges, we find that the pooling of seller IDs by unrelated publishers is also widespread (\Cref{sec:pooling:pooling}). Misinformation websites, which violate the publisher content policies of many AdXs, are able to monetize their ad inventory through these pools. 
In fact, we find that in many cases they are able to leverage the authorized programs of the same AdXs whose policies they violate.
%

%%
% Unused Seller ID         & 2524 & 54 \\ \hline
% Sellers.json Unavailable & 1951 & 9  \\ \hline
% Domain Unavailable       & 135  & 12  \\ \hline
% Confidential Seller      & 86   & 13 \\ \hline
\section{Brand Analysis} 
\label{sec:brands}
In this section, we analyze the display ads loaded on misinformation websites to identify the advertisers/brands that end up buying their ad inventory. 

\begin{figure}[!t]
\centering
\includegraphics[width=.98\columnwidth]{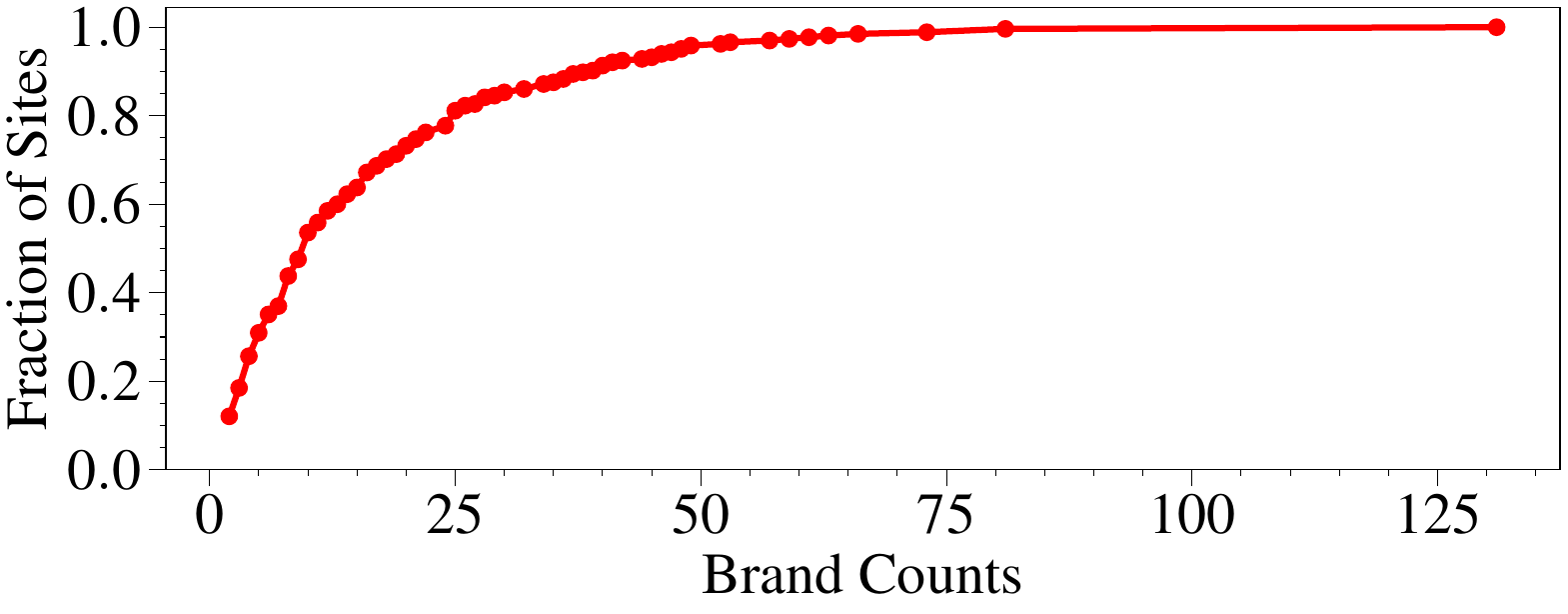}
\caption{Cumulative distribution of the number of distinct brands across different misinformation websites}
\label{fig:cdf_distinct_brands_across_mdoms_a}
\bigvspace
\end{figure}

\para{Data collection.}
We curate \databrands by crawling each of the 669 misinformation websites ten times as discussed in \Cref{sec:data:methodology}. 
We are able to collect a total of 4,246 ads belonging to 2,068 distinct brands.
Figure~\ref{fig:cdf_distinct_brands_across_mdoms_a} plots the distribution of the number of distinct brands across misinformation websites.
We find that a non-trivial fraction of misinformation websites are able to get ads from tens of distinct brands. 
Specifically, 23 misinformation websites have ads from at least 41 distinct brands each while 142 misinformation websites have ads from at most 10 distinct brands each.

\begin{figure}[!t]
\bigvspace
\centering
\includegraphics[width=\linewidth]{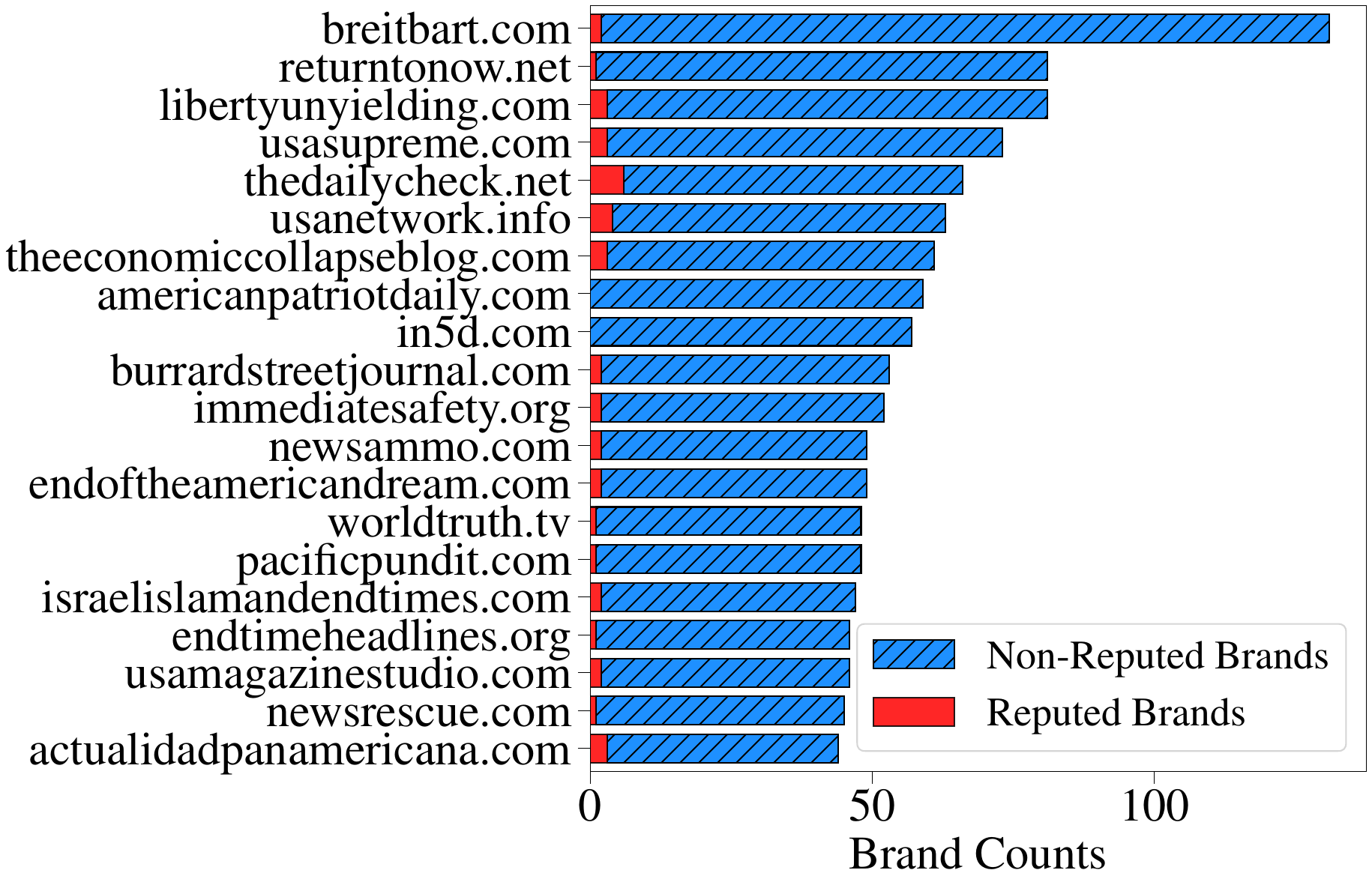}
\caption{Distribution of reputable and non-reputable brands among the top-20 misinformation websites with the highest number of distinct brands advertising on their website.}
\label{fig:bar_rep_nonrep_top20_misinfo_with_max_brand}
\bigvspace
\end{figure}

\para{Reputable brand classification and prevalence.}
To assess whether these ads are from reputable brands, we use their Tranco ranks as a rough proxy for their reputation. 
Specifically, we classify brands with top-1K Tranco ranking as ``reputable''. 
Figure \ref{fig:bar_rep_nonrep_top20_misinfo_with_max_brand} shows the number of distinct reputable and non-reputable brands across top-20 misinformation websites that contain ads from the highest distinct brands.
Perhaps surprisingly, we find that Breitbart -- a well-known misinformation website -- is able to attract ads from the highest number of distinct brands. 
The two reputable brands with ads on Breitbart include {Forbes} and {GoDaddy}.
In total, we observe ads from 55 reputable brands including Forbes, GoDaddy, Harvard, Intel, Microsoft, Nike, Samsung, Tumblr, Yahoo!, Verizon, and Wayfair.
We note that these top-20 misinformation websites tend to have more ads from reputable brands on average as compared to the remaining misinformation websites. 
Specifically, the average number of reputable brands is 2.05 for the top-20 misinformation websites in Figure \ref{fig:bar_rep_nonrep_top20_misinfo_with_max_brand} and 0.78 for the remaining misinformation websites.

\para{Correlation between ad inventory misrepresentation and number of brands.}
Next, we investigate whether the misrepresentation of ad inventory by misinformation websites impacts their ability to sell their ad inventory.

\begin{figure}[!t]
\centering
\includegraphics[width=.98\columnwidth]{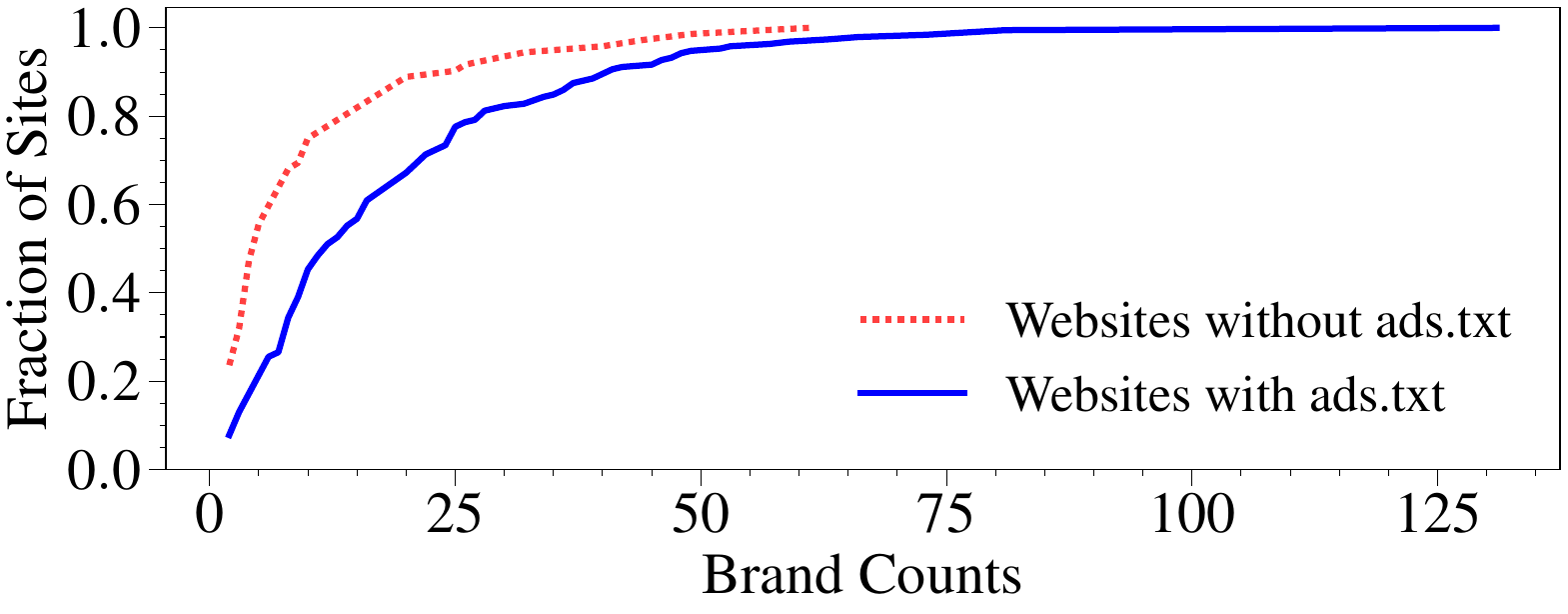}
\caption{Cumulative distribution of the number of distinct brands on misinformation websites with or without \ads}
\label{fig:cdf_distinct_brands_across_mdoms_b}
\bigvspace
\end{figure}

Figure~\ref{fig:cdf_distinct_brands_across_mdoms_b} plots the distribution of the distinct brand counts of all the brands advertising on misinformation websites with/without \texttt{ads.txt}.
Note that we are looking for the existence of \texttt{ads.txt}.\footnote{{The mere existence of \texttt{ads.txt} does not guarantee the veracity of its content. A misinformative publisher could have misrepresented \ads entries to bypass brand checks. Advertisers are recommended by IAB to perform \ads checks against the data observed in the bid requests before making a bid.}}
We find that misinformation websites with \texttt{ads.txt} are able to attract ads from twice as many brands on an average as compared to the websites without \texttt{ads.txt}. 
We conclude that some brands do avoid advertising on misinformation websites without \texttt{ads.txt}.

\begin{figure}[!t]
  \centering
  \includegraphics[width=0.9\linewidth]{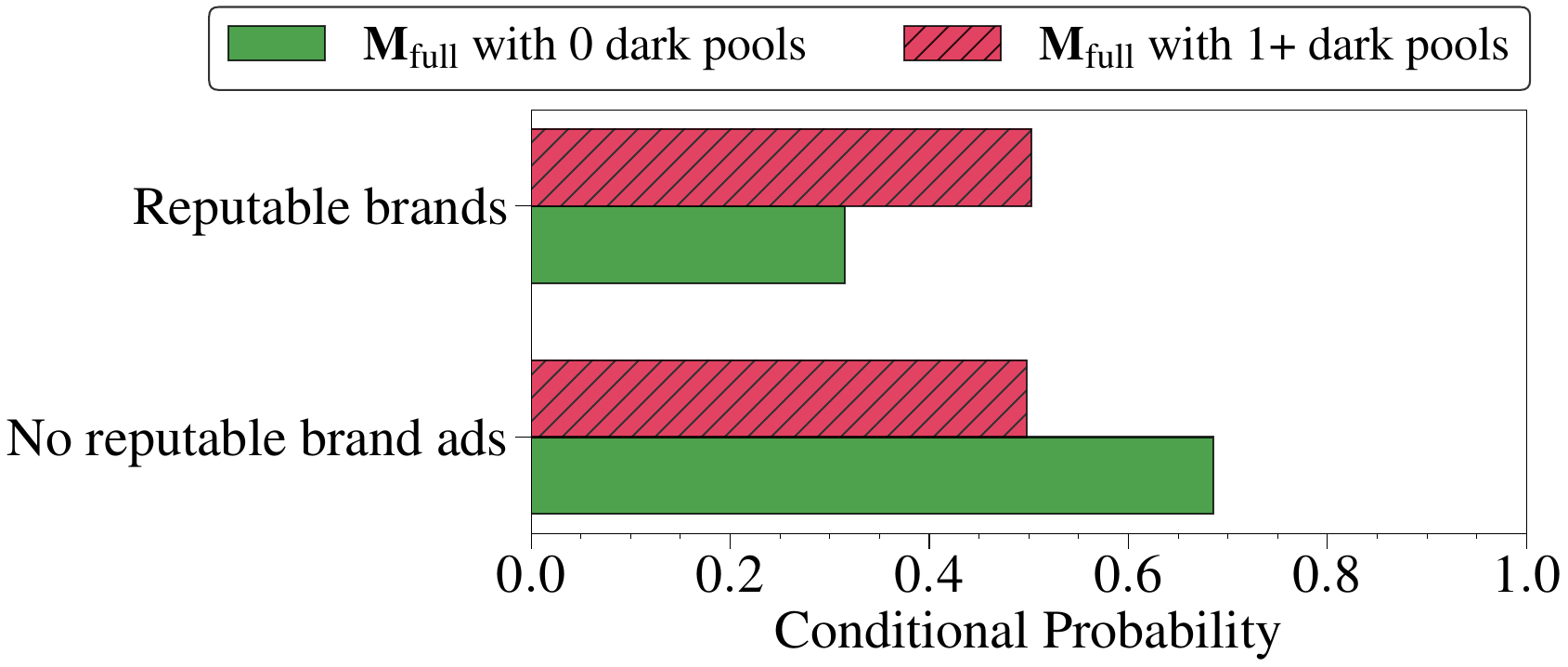}
  \caption{Conditional probabilities of ads from reputable brands in presence or absence of dark pooling}
  \label{fig:conditional_probabilities}
  \bigvspace
\end{figure}

Figure~\ref{fig:conditional_probabilities} plots the conditional probabilities of observing reputable brands across misinformation websites with/without dark pools.
We find that more than half of the misinformation websites part of one or more dark pools get ads from reputable brands. 
In contrast, less than one-third of the misinformation websites part of no dark pools get ads from reputable brands.
This nearly 20\% difference in the conditional probability shows that dark pooling significantly increases the chances of ads from reputable brands ending up on misinformation websites.

\para{Brand disclosures.}
It is reasonable to assume that reputable brands generally do not want to advertise on misinformation websites \cite{AdvertisersDemonetizedBreitbart, IASBrandFavorability, IASConsumerPerception}. 
Taking the example of Breitbart, there is ample evidence that reputable brands did not want their ads shown on Breitbart \cite{AdvertisersDemonetizedBreitbart, SunrunBreitbart, SageNAmericaBreitbart}.
DSPs and AdXs typically provide brand safety features \cite{bellman2018brand} to help brands avoid buying the ad inventory of low-quality websites. 
Brand safety features allow brands to block unwanted ad inventory through a block list of domains or seller IDs \cite{PubIDBlocking}.
One would expect that reputable brands would attempt to avoid buying the ad inventory of misinformation websites through these brand safety features. 
Since brand safety is not externally measurable, we conduct individualized disclosures to these 55 reputable brands and specifically ask them (a) whether they want their ads on misinformation websites or not and (b) whether they employ brand safety features to this end. 

% At least 15 out of the 55 reputable brands in our dataset (Forbes, GoDaddy, Spotify, Hulu, etc) had previously acknowledged \cite{checkmyads,AdvertisersDemonetizedBreitbart} that their ads on misinformation websites were unintended and/or in violation of the intended brand safety controls.
%
% In fact, this has become a major point of emphasis in recent brand safety initiatives \cite{AdvertisersDemonetizedBreitbart, IASBrandFavorability, IASConsumerPerception}.

To perform disclosures, we first attempted to find advertising-related email addresses for each reputable brand from their website. 
If we were unsuccessful, we included generic email addresses from their ``About Us'' and ``Contact Us'' pages. 
In our disclosures, we listed the misinformation websites where the ads of the reputable brand were observed. 
We included full-page screenshots showing the brand's ad creative on the misinformation website as well as the full HTTP Archive (HAR) recording of the network traffic. 
We also asked them whether they were aware of or intended to have their ads on the misinformation websites and whether/which brand safety service they used.

We received responses from 11 reputable brands. 
8 brands confirmed that they were unaware and did not intend to advertise on these misinformation websites. 
For example, one brand responded that ``\textit{We don’t advertise on the site. It was an unintentional oversight related to automated advertising and the ad was immediately pulled when discovered. We always aim to advertise on sites that are aligned with our mission and values and we apologize if this upset any of our customers.}'' 
Another brand mentioned that ``\textit{We will not want to see our ads on misinformation websites.}'' 
%
% We had back-and-forth conversations with 6 brands and helped them further fix the issue. 
%
Regarding the deployment of brand safety features, we received confirmation from 4 brands that they indeed used a brand safety service but it did not adequately detect or prevent their ad from appearing on the misinformation website. 
One brand told us that it used Google Display Network's built-in brand-safety measure while two brands employed the brand safety service provided by Integral Ad Science (IAS). 
One brand told us that ``\textit{the misinformation website disclosed by you [...] is neither present in the logs provided to us by our DSP partner nor is flagged by IAS. We think that it is being misplaced on the misinformation website due to dark pooling.}''
Another brand told us that ``\textit{About the brand-safety service, please understand we are not able to tell any detail}'' presumably due to a confidentiality agreement.

% Furthermore, it has been reported that 82\% of advertisers use some form of brand-safety technology~\cite{BrandSafetyUsage}. 
% Thus, it is reasonable to assume that most of the cases observed in our data are not because the brands do not use brand-safety at all.
%

\begin{comment}
    - 11 reputable brands that replied: Forbes, Financial Times, GoDaddy, Harvard, Intel, Mozilla, Spotify, AOL, Yahoo, Cloudflare, T-mobile
    - 8/11 brands that replied positively: Forbes, GoDaddy, Harvard, Intel, AOL, Yahoo, Cloudflare, T-mobile
    - 6/8 brands that we followed-up with: Forbes, GoDaddy, Harvard, Intel, AOL, Yahoo
    - 2/8 brands that we virtually met: Cloudflare, T-mobile
    - 4/8 brands that replied about brand-safety: Harvard, Intel, Cloudflare, T-mobile
\end{comment}

%%
\para{Takeaways.}
In summary, our results show that the misrepresentation of ad inventory by misinformation websites seems to be correlated with their ability to monetize their ad inventory through reputable brands. 
We found that the ad inventory of misinformation websites that use dark pooling is more likely to be bought by reputable brands. 
The limited responses from reputable brands suggest that they do not want to advertise on misinformation websites and employ brand safety features to this end.

\begin{comment}
\footnote{To concretely study brand safety, we would need proprietary data from ad exchanges or brand safety services. 
%
Since this data is not available to us, we are unable to provide individualized evidence whether reputable brands not want their ads on misinformation websites and that their brand safety efforts are undermined due to the aforementioned ad inventory misrepresentations. 
%
That said, it is well-known in the online advertising industry that reputable brands generally do not want to advertise on misinformation websites. 
%
In fact, this has become a major point of emphasis in recent brand safety initiatives \cite{AdvertisersDemonetizedBreitbart, IASBrandFavorability, IASConsumerPerception}.
%
Taking the example of Breitbart, there is ample evidence that brands were unaware and objected to having their ads shown on Breitbart \cite{AdvertisersDemonetizedBreitbart, SunrunBreitbart, SageNAmericaBreitbart}.
}
\end{comment}

\section{Related Work} \label{sec:related}

\para{Examining the online advertising ecosystem.}
In recent years, there have been many research efforts to bring transparency to the mechanisms of online advertising. 
A large number of these have focused on studying personal data collection and sharing to deliver personalized ads~\cite{englehardt2016online, razaghpanah2018apps, bashir2016tracing, cahn2016empirical, vekaria2021differential, falahrastegar2014rise}. 
Our work instead focuses on the prevalence of inventory fraud, pooling, and its impact on brands. 

\parait{Inventory fraud.}
There have been a few measurements related to \ads standard and related inventory fraud since its introduction. However, no work has focused on \sellers or the ad-fraud that emerges by the combined failure of \ads and \sellers.
In 2019, Bashir \etal \cite{bashir2019longitudinal} gathered and conducted a longitudinal analysis of \ads files. They found that these files were riddled with syntactic errors and inconsistencies that made them difficult to process in an automated fashion. 
Tingleff \cite{tingleff2019three} and Pastor \etal \cite{pastor2020establishing} highlighted flaws of the \ads standard 
that undermines its effectiveness in preventing ad fraud, albeit without measurements to support their hypotheses. Some of these identified flaws are, however, supported by measurements from Papadogiannakis \etal \cite{papadogiannakis2023funds}.
These findings, suggesting that the \ads standard is not effectively enforced, are corroborated by our study.
Our work complements these efforts by undertaking a measurement study of both the standards of \ads and \sellers for the first time to measure inventory fraud as well as prevalence of pooling, which allows low-quality publishers to launder their ad inventory. 

\parait{Brand safety.}
There have been many studies that have highlighted the impact of ads (and the websites on which they appear) on the reputation of a brand~\cite{bellman2018brand, shehu2021risk, lee2021spillover, bishop2021influencer}. 
In fact, several activist efforts have successfully leveraged brand
safety concerns to demonetize misinformation. Notable among these are the efforts of Check My Ads and Sleeping Giants~\cite{ferraz2021sleeping}, who successfully used public campaigns to pressurize 820 brands to add Breitbart News' domain to their advertising block lists. 
Our cataloging of brands found on known misinformation websites can supplement these ad-hoc efforts and increase pressure on ad-tech to enforce its own \ads and \sellers standards more effectively. 
Other work has focused on measuring or improving the effectiveness of mechanisms for identifying `brand safe' web content. 
Most recently, Vo \etal \cite{vo2020unsafe} built an image-based brand-safety classifier to prevent ad placement on inappropriate pages. Numerous products from major ad-tech firms such as DoubleVerify \cite{brandsafety-doubleverify},
Integral Ad Science \cite{brandsafety-ias}, and Oracle \cite{brandsafety-moat} have also recently started promoting their `brand safety' features. 
\para{Funding infrastructure of misinformation.}
Ours is not the first work to consider the role of the online advertising ecosystem in funding misinformation.
In fact, it has been known for several years that online advertising provides the primary revenue stream for misinformation websites~\cite{kshetri2017economics, article1, article2, article3, article4}.
Han \etal \cite{han2022infrastructure}, in their study, focused on network infrastructure, also explored the revenue streams on misinformation websites and identified disproportionately high reliance on advertising and consumer donations. 

Bozarth \etal \cite{bozarth2020market} showed that although there is a unique ecosystem of `risky' AdXs that partner with publishers of misinformation, there is also a heavy presence of mainstream AdXs (\eg Google) in the misinformation ecosystem. 

Braun \& Eklund \cite{braun2019fake} take a qualitative approach to understand the role of the advertising ecosystem in increasing revenues of misinformation and the dismantling of traditional journalism. 
Their work, along with numerous others~\cite{timmer2016fighting, tambini2017fake, vasu2018fake}, has highlighted the need for additional 
transparency to realize the promise of market-based strategies to curb funding of misinformation.

Considering another angle, several studies have also examined how deceptive ads are used to promote and fund harmful products~\cite{mejova2020advertisers, jamison2020vaccine, boudewyns2021two} and ideologies~\cite{zeng2020bad, zeng2021polls, chen2022misleading}.

At a high-level, our work complements all these efforts to better understand how the misinformation ecosystem is funded by online advertising by uncovering and analyzing the exploitation of specific advertising-related vulnerabilities such as pooling and relationship misrepresentations by the misinformation ecosystem.

\section{Concluding Remarks}
\label{sec: conclusion}
Our work shows how the opacity of the ad-tech supply chain is exploited by misinformation publishers to monetize their ad inventory. 
Through our measurements, we demonstrate a widespread lack of compliance with the IAB's \ads and \sellers standards, ad inventory pooling by misinformation publishers, and reputed brands who end up buying this ad inventory of misinformation publishers. 
Taken all together, our results point to specific gaps that need to be further explored by the ad-tech and security research communities.

\para{Trust delegation in advertising partner programs.}
% Blurb for delegation of trust and trust vetting in PKI, play stores, etc. and parallels to this paper.
One of our key findings is that a small number of ad exchanges are responsible for a majority of dark pooling. In many cases, we see evidence that this dark pooling is achieved through the use of legitimate partner programs made available by ad exchanges (\eg Google's MCM partner program \cite{GoogleMCM}).
These programs serve an important purpose --- to help reduce the management burdens on small publishers. However, as we see in our study, this expanded access facilitated via advertising partners results in new vulnerabilities. Specifically, publishers who are in clear violation of the policies set by an ad exchange are still able to obtain seller IDs issued by the exchange through their partners.
One perspective of this problem is that there is a fundamental breakdown of trust delegation --- \ie partners are delegated the rights to assign and manage seller IDs on behalf of exchanges, but without being properly delegated the responsibilities for vetting publishers and verifying their compliance with ad exchange policies.
While this work is the first to uncover this delegation of trust in the form of verification responsibilities in the ad-tech ecosystem, it is not new to the security community.
Indeed, this type of trust delegation is a central theme in the public key infrastructure \cite{chuat2020sok}, app stores \cite{lin2021longitudinal}, and other domains. 
From these prior efforts to delegate verification responsibilities, it is clear that success is only possible with effective mechanisms to monitor compliance and revoke delegated trust. A key difference from prior efforts, however, is that it is not publicly known how these trust delegation processes work within specific ad exchanges.
Without public documentation of these processes or research studies that uncover them, we anticipate that identifying weaknesses and causes for failure will remain an open challenge.

\para{Supply chain transparency and compliance with industry standards.}
The programmatic advertising supply chain is complex because of the large number of entities involved between publishers and advertisers. Further complicating matters, our study shows that these entities are frequently out of compliance with even basic standards such as \ads and \sellers.
In fact, many of the concerning findings of our work could be addressed if advertisers were able to trace the provenance of ad inventory using the existing `Supply Chain Object' (SCO) ad-tech standard.
Unfortunately, our analysis of SCOs in $\S$\ref{appendix: supply chain object} shows that less than a quarter of bid requests actually include the SCO. Further, even when the SCO is included in bid requests, they are often incomplete and missing information would make end-to-end verification of the supply chain difficult.
We further find that even major ad exchanges implement digital advertising standards in a way that hinders external independent audits. 
Notably, Google's widespread use of confidential \sellers entries \cite{PubIDBlocking} makes it challenging to identify Google AdX's partners who are not doing adequate compliance verification for the publishers whose inventory they list. 
IAB has recently released new and updated digital advertising standards \cite{ads-cert,adstxt-updated} to improve end-to-end validation of the ad-tech supply chain. 
However, these are not widely adopted yet. 
Therefore, in its current state, to mitigate ad fraud and reduce ad-tech's inadvertent funding of misinformation, it is crucial that adoption and compliance with new and existing digital advertising standards such as SCO, \ads, and \sellers improve.
However, a key challenge is the absence of incentives for achieving compliance with these standards. It remains to be seen if recent US regulatory efforts will improve compliance. Notably, the Digital Services Oversight and Safety Act (DSOSA) \cite{dsosa} and Advertising Middlemen Endangering Rigorous Internet Competition Accountability Act (AMERICA) \cite{americaact} introduce new requirements related to online advertising transparency.
In addition, we are currently engaged in conversations with members of US Congress seeking to draft additional legislation specifically to strengthen compliance with ad-tech industry standards, improve transparency around the ad-tech supply chain, and mitigate ad fraud.

\para{Effective notification and vulnerability reporting mechanisms.}
The ad-tech industry is currently lacking mechanisms through which supply chain vulnerabilities may be reported. 
This absence has resulted in several community-organized efforts such as the Check My Ads Institute \cite{checkmyads} that monitor ads on misinformation websites and use social media to report on the brands or ad-tech loopholes that fund these publishers. 
While these efforts have been successful at mitigating some of the harms from the opacity of the supply chain, they are not systematic reports and rely on amplification via social media in order to reach their intended targets. Further, like our study, they generally focus on specific harms caused by the opacity of ad-tech (e.g., funding of misinformation). 
There is a need to develop more generalized and systematic mechanisms for reporting supply chain vulnerabilities and non-compliance with existing industry standards. 
%
%We are currently collaborating with the Check My Ads Institute to disclose the supply chain vulnerabilities identified in this paper to relevant stakeholders.

For reproducibility and to foster follow-up research, our dataset is available at
\url{https://osf.io/hxfkw/?view_only=bda006ebbd7d4ec2be869cbb198c6bd5}

%% Initial Outline Plan:
% - [MCM vetting]
% - [Supply Chain Object]
% - [ads.cert: Cryptographic]
% - [Brand-safety: Post-bid (if impression is won then the DSP/advertiser can verify that they have been duped)]

\section*{Acknowledgment} 
This work is supported in part by the National Science Foundation under grant numbers 2103439, 2103038, and 2138139.
We want to thank the anonymous shepherd and reviewers for their constructive feedback that helped improve the work.

% \newpage
\bibliographystyle{unsrt}
\bibliography{mybib}

% % The Meta-Review should at least start on a new column

% Use \appendices and not \appendix due to IEEEtran.cls quirks

\vfill\eject
\appendices % if not used earlier

\section{Longitudinal Analysis of \sellers}
\label{appendix: longitudinal}
Various campaigns have highlighted the role of AdXs in monetizing the misinformation ecosystem, pressuring them to remove their
support for these domains \cite{checkmyads}. 
To understand the effectiveness of these campaigns, we monitored changes to the \sellers files present in our \datastatic{.} dataset for a three-month period (from Oct'21 to Feb'22).
Of the 470 AdXs found to support misinformation websites (by listing them as publishers) on October 2021, 39 (8.3\%) AdXs delisted at least one
misinformation website by February 2022. 

% It is important to understand whether the AdXs are de-listing the misinformation content from their \texttt{sellers.json} files. 
%
Bashir et. al.~\cite{bashir2019longitudinal} performed this analysis on \texttt{ads.txt} of Alexa Top-100K websites in their work. 
However, our study is on misinformation websites, whose \texttt{ads.txt} should not be trusted. 
Hence, we perform this analysis on \texttt{sellers.json} files of trusted AdXs.

We observed 470 \texttt{sellers.json} supporting at least one misinformation website as per October's crawl -- 46 of which support 10 or more misinformation outlets. The ones that support the highest misinformation websites are \textit{revcontent.com} (204), \textit{liveintent.com} (56), \textit{outbrain.com} (56), \textit{pixfuture.com} (39), and \textit{lijit.com} (now part of \textit{Sovrn}) (30). 
From Oct'21 to Feb'22, only 39 AdXs de-list at least 1 misinformation website, while 53 \texttt{sellers.json} include at least 1 misinformation website in their files. 
Table~\ref{tab:longitudinal_analysis} shows the top AdXs and their longitudinal support for the misinformation websites in their \texttt{sellers.json}. 

\begin{table}[!hbt]
\centering
\small
\begin{tabular}{lcccc}
\hline
\multicolumn{1}{c}{\multirow{2}{*}{\textbf{Ad exchange}}} \\[-1em] & \multicolumn{4}{c}{\textbf{Misinformation Website Counts}}             \\ \cline{2-5} 
\multicolumn{1}{c}{}                                     \\[-1em] & \textbf{Oct'21} & \textbf{Feb'22} & \textbf{Added} & \textbf{Dropped} \\ \hline
revcontent.com                                            & 204             & 73              & 2              & 133              \\ \hline
outbrain.com                                              & 56              & 35              & 0              & 21               \\ \hline
9mediaonline.com                                          & 20              & 1               & 0              & 20               \\ \hline
stitchvideo.tv                                            & 14              & 1               & 0              & 13               \\ \hline
adtelligent.com                                           & 26              & 28              & 13             & 11               \\ \hline
infolinks.com                                             & 23              & 14              & 2              & 11               \\ \hline
publisherdesk.com                                         & 14              & 3               & 0              & 11               \\ \hline
mgid.com                                                  & 20              & 32              & 13             & 1                \\ \hline
nextmillennium.io                                         & 7               & 9               & 3              & 1                \\ \hline
vidazoo.com                                               & 5               & 8               & 3              & 0                \\ \hline
pixfuture.com                                             & 39              & 41              & 2              & 0                \\ \hline
lijit.com                                                 & 30              & 30              & 0              & 0                \\ \hline
\end{tabular}
\caption{AdXs that add and drop the most misinformation websites from their \texttt{sellers.json} between Oct'21 and Feb'22. The table is arranged in descending order of the dropped counts.}
\label{tab:longitudinal_analysis}
\smallvspace
\end{table}

Upon further investigation of RevContent, we observed that it dropped  $\sim$87\% of the total publisher domains from their \texttt{sellers.json} in mid-December 2021 (Oct'21: 4727 domains to Feb'22: 621 domains) and we speculate that their primary aim might not have been to drop misinformation websites, but they ended up de-listing a few of misinformation websites too as a result of their bulk drop. 
There has always been a constant peer-pressure and criticism from activists (e.g., \cite{checkmyads}) forcing RevContent to remove their support for misinformation websites. 
There were active discussions on social media speculating RevContent's intent behind this massive drop. 
However, RevContent did this silently and never publicly addressed this action. 
%
% It is surprising that eventhough \textit{RevContent} dropped the maximum number of misinformation websites from its \texttt{sellers.json}, it still potentially funds the highest online misinformation. 
Even after the drop, {RevContent} still potentially funds the most misinformation websites in our data.
Other than RevContent, other AdXs that continued their support for the highest misinformation websites in Feb'22 are \textit{LiveIntent} (56), \textit{Pixfuture} (41), \textit{Outbrain} (35), and \textit{MGID} (32).

Additionally, the misinformation outlets which were added by the most AdXs are \textit{rearfront.com}, \textit{vidmax.com}, and \textit{thetruereporter.com}. 
The former 2 outlets are agents of spreading viral and misleading content, while the latter publishes politicized news, commentary and analysis. 
These were added by 6 different AdXs. 
Similarly, \textit{lifezette.com}, \textit{waynedupree.com}, and \textit{news18.co} were dropped by 6, 6, and 5 AdXs respectively.

\section{Supply Chain Object Analysis}
\label{appendix: supply chain object}
If adopted and implemented correctly, Supply Chain Objects (SCOs) can aid overall validation and provide transparency into all the entities involved in (re-)selling of a particular ad-inventory. 
In absence of SCOs, a buyer has visibility into only the immediate upstream seller but not the entire path of (re-)sellers that were involved before the upstream seller. 
It is the job of each seller to append its seller object in the existing SCO and forward the bid request further. 
A buyer extracts the SCO object from the bid request and parses the list of all seller nodes represented by key {\tt nodes}. %(which contains a list of dictionaries). 
%
% Lower the index of a node in this list, the more older the seller.
Higher the index of a node in this list, the more recent the seller. 
%
% When SSP forwards the bid request corresponding to the ad request received from the publisher, in the node dictionary it appends its website (represented by the key {\tt asi}) and account identifier for the given publisher in its network (represented by the key {\tt sid}).
When an AdX forwards the bid request for a publisher, it associates the publisher with dictionary key {\tt asi} and the account identifier for that publisher in its network with the key {\tt sid}.

In order to analyze the adoption and correctness of SCOs in our data, we use our custom SCO parser (based on the IAB guidelines) on all the bid requests captured in the \datacrawls dataset.
Despite SCOs being introduced by IAB since July 2019, only 20.5\% (3796) bid requests have included SCOs, all of which comprised only a single seller node. 
To verify the correctness of SCOs, we extracted {\tt sid} and {\tt asi} associated with the seller node and performed {\tt sid} lookup in the \sellers file of the {\tt asi} to obtain the upstream seller domain with which the ad-inventory is associated as per the SCO. 
Next, we checked if this website domain matched the actual website's domain on which the current bid request was captured during the dynamic crawl. Let's call this boolean result -- A. 
%
% We, also generated a dynamic path based on the SCO object as follows: upstream website $\rightarrow$ asi seller $\rightarrow$ domain of the ad-request. 
% %
% We obtain the ground truth from the sellers.json, using which we also generated all 3-hop static paths for each misinformation website in our dataset. 
% %
% Next, we checked each of the 3796 dynamic paths among all the static paths of the associated misinformation website. 
We also validated all 3796 SCO-based paths (upstream website $\rightarrow$ {\tt asi} seller $\rightarrow$ ad-request domain) against 3-hop static paths involving each misinformation website generated from the \sellers files.
Let's call this boolean result -- B. 
The cases where A and B were True are cases where we could verify the correctness of the SCOs. 
The rest cases were SCO misrepresentations. 
We observed only 18.94\% (719) of 3796 bid requests with correct implementation of SCOs.

\newpage

\end{document}